\documentclass[pre,twocolumn,superscriptaddress]{revtex4-2}
\usepackage{amsmath,amssymb,dsfont,mathtools,booktabs,longtable,color,soul}
\usepackage[utf8]{inputenc}
\usepackage[T1]{fontenc}
\usepackage[english]{babel}

\newcommand{\lp}{\left(}
\newcommand{\rp}{\right)}
\newcommand{\ls}{\left[}
\newcommand{\rs}{\right]}
\newcommand{\lbr}{\left\lbrace}
\newcommand{\rbr}{\right\rbrace}
\newcommand{\TP}{\mathrm{TP}}
\newcommand{\TN}{\mathrm{TN}}
\newcommand{\FP}{\mathrm{FP}}
\newcommand{\FN}{\mathrm{FN}}

\begin{document}
\title{Reconstructing simplicial complexes from evolutionary games}
\author{Yin-Jie Ma}
    \affiliation{School of Business, East China University of Science and Technology, Shanghai, China}
    \affiliation{Research Center for Econophysics, East China University of Science and Technology, Shanghai, China}
    \affiliation{CNR - Institute of Complex Systems, Via Madonna del Piano 10, I-50019, Sesto Fiorentino, Italy}
\author{Zhi-Qiang Jiang}\email{Corresponding author: zqjiang@ecust.edu.cn}
    \affiliation{School of Business, East China University of Science and Technology, Shanghai, China}
    \affiliation{Research Center for Econophysics, East China University of Science and Technology, Shanghai, China}
\author{Fanshu Fang}
    \affiliation{CNR - Institute of Complex Systems, Via Madonna del Piano 10, I-50019, Sesto Fiorentino, Italy}
    \affiliation{College of Economics and Management, Nanjing University of Aeronautics and Astronautics, Nanjing, China}
\author{Charo I. \surname{del Genio}} \email{Corresponding author: charo.delgenio@trakia-uni.bg}
    \affiliation{Institute of Smart Agriculture for Safe and Functional Foods and Supplements, Trakia University, Stara Zagora 6000, Bulgaria}
    \affiliation{Research Institute of Interdisciplinary Intelligent Science, Ningbo University of Technology, 315104, Ningbo, China}
\author{Stefano Boccaletti}
    \affiliation{CNR - Institute of Complex Systems, Via Madonna del Piano 10, I-50019, Sesto Fiorentino, Italy}
    \affiliation{Research Institute of Interdisciplinary Intelligent Science, Ningbo University of Technology, 315104, Ningbo, China}
    \affiliation{Sino-Europe Complexity Science Center, School of Mathematics, North University of China, 030051, Taiyuan, China}

\date{\today}

\begin{abstract}
In distributed systems, knowledge of the network structure of the connections among the unitary components
is often a requirement for an accurate prediction of the emerging collective dynamics. However, in many real-world situations, 
one has, at best, access to partial connectivity data, and therefore the entire graph structure needs to be
reconstructed from a limited number of observations of the dynamical processes that take place on it. While existing studies 
predominantly focused on reconstructing traditional pairwise networks, higher-order interactions remain largely unexplored. 
Here, we introduce three methods to reconstruct a simplicial complex structure of connection from observations of evolutionary games
that take place on it, and demonstrate their high accuracy and excellent overall performance in synthetic  and empirical
complexes. The methods have different requirements and different complexity, thereby constituting a series of approaches from which one can
pick the most appropriate one given the specific circumstances of the application under study.
\end{abstract}

\maketitle

\section{Introduction}
Complex networks provide a powerful framework to study
a broad range of systems~\cite{Alb02,Boc06,Boc14}. Traditional
networks consist of a number of links, called edges, between
pairs of discrete elements, called nodes. Their use has
proved highly successful in numerous applications, which
include determining the functional modules of the human
brain~\cite{Pap14}, describing the dynamics of gene expression~\cite{Min13},
detecting the presence of communities in large data sets~\cite{Tre15,Bot16,Bot17},
modelling the emergence of strategy in social interactions~\cite{Hil17,Per17,Li18,Wan18,Sch21},
and suggesting the formulation of new antimicrobials~\cite{Con20,Con22}.
Moreover, they have been a fundamental tool in investigating
the properties of dynamical processes such as epidemic
and information spreading~\cite{del13,Zha16,Zha22}, evolutionary
games~\cite{del11}, and synchronization of coupled oscillators~\cite{del15,del16,del22}.
However, a limitation of the original network paradigm
has recently come to light, with the realization that
many complex systems feature multi-body interactions,
whereby macroscopic effects result from collective contributions,
each involving three or more nodes~\cite{Boc23}.

Accounting for the presence of higher-order interactions
induces a profound change in the mathematical structure
of the networks. In fact, while traditional networks are
usually described by simple graphs, higher-order networks
are best represented as hypergraphs, which are collections
of edges that can link any number of nodes. A special case
occurs when the presence of a collective interaction between
the members of a set of nodes induces additional interactions
amongst all possible smaller subsets of the same nodes.
The resulting structure is then a particular case of a hypergraph,
which takes the name of \emph{simplicial complex}. Numerous
theoretical advances in the study of hypergraphs and simplicial
complexes have been accomplished over the last decade. However,
virtually all network methods rely on the knowledge
of the global structure, which is often not accessible when
investigating real-world systems. Thus, researchers are
forced to find ways to infer or reconstruct the unknown
links from limited observations.

\begin{figure*}[t]
\centering
\includegraphics[width=0.65\textwidth]{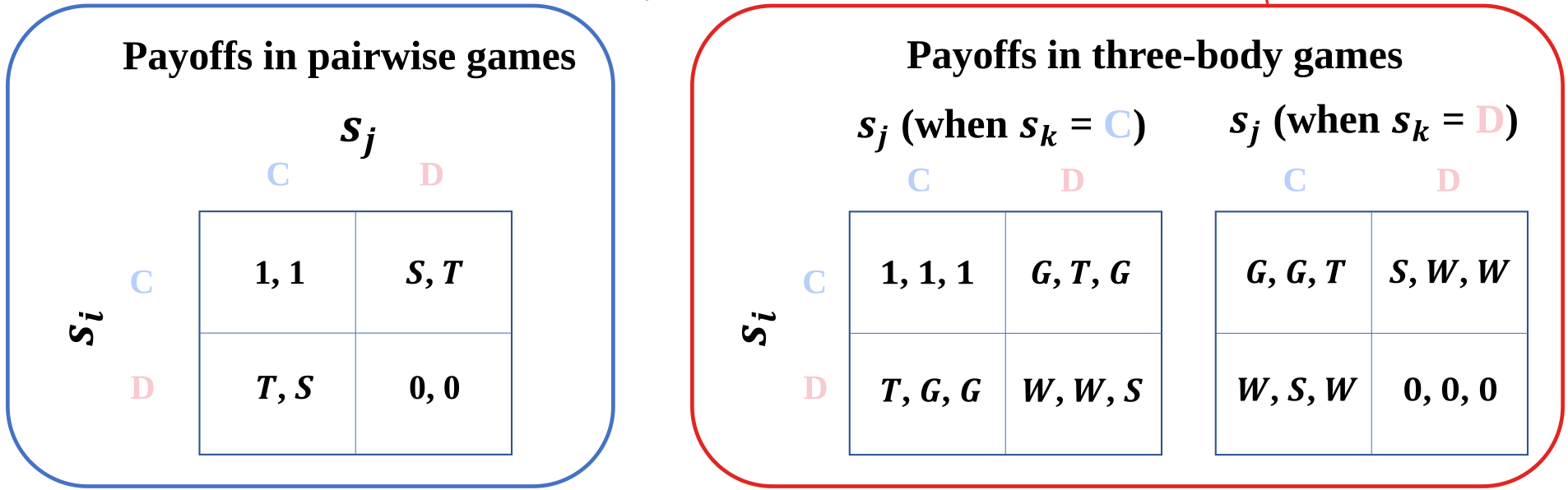}
\caption{\textbf{Payoff matrices for evolutionary games.} If the
players' strategy~$s$ is to cooperate~(C) or defect~(D) all at
the same time, they all receive a unitary reward or nothing, respectively.
If a player is the only one to cooperate or defect, it receives
a sucker's payoff~$S$ or a temptation payoff~$T$. In three-body
games, two players who cooperate against a single defector receive
a payoff~$G$, whereas two players who defect against a single cooperator
receive a payoff~$W$.}\label{Fig:GraphicIllustration}
\end{figure*}
Because of this, the inverse problem of determining the structure of a network
from partial data has sparked a notable amount of interest, and many methods to
achieve this goal have been developed. Depending on the nature of the approach
they take, these can be roughly ascribed to two types, namely the statistical
ones~\cite{Her08,Che17,Run18,Wan22,Xu23}, which aim to detect the edges in the
network using inference methods, and the optimization ones~\cite{Wan11,Han20,Hua20,Dai21,Ma23},
which turn the global task into a set of sub-problems of convex optimization with
linear constraints. Within this latter class, compressive sensing~ has gained
a great popularity during the last decade~\cite{Squ18}. The main idea behind the technique
is to take a series of observations, whose nature depends on the specific kind
of system being reconstructed, and use it to derive a set of underdetermined linear
equations. In the assumption that the network is sparse, which is almost always
true in real-world situations~\cite{del11_2}, one can then solve the system and
produce local estimates of the adjacency matrix, which encodes the full connectivity
of the network. Note that, despite the effectiveness of compressive sensing, care
must be taken to ensure that the results obtained are consistent. For example, in undirected
networks, one must ensure that the resulting adjacency matrix is symmetric, which
can be achieved via the inclusion of latent constraints~\cite{Hua20}. It is also
worth noticing that such constraints may go beyond the symmetry ones when higher-order
interactions are considered~\cite{Mal24}, and their number grows quickly with the
order of the edges. Thus, developing an efficient reconstruction method for higher-order
networks is a serious challenge.

In this article, we propose a novel framework for reconstructing
simplicial complexes that effectively captures higher-order interactions
in dynamic real-world settings, where equilibrium is often impractical
due to the fluctuating nature of social and strategic interactions. A
key distinction of our approach is the utilization of transient-phase
data from evolutionary games, rather than relying on equilibrium states
or non-equilibrium steady states. Unlike the latter, transient phases
provide richer dynamical information, encoding the system’s pathway
toward equilibrium and offering a more effective means of inferring
higher-order structures. Transient dynamics frequently emerge in
evolutionary games, such as coordination games on complex networks~\cite{Chi24}, prisoner’s dilemma with adaptive strategies~\cite{Han22}, and
public goods games with reputation-driven interactions~\cite{McI13}, which reinforce the relevance of our approach. Our framework is based on
compressive sensing techniques and includes a point-by-point estimator~(PBP), a
global estimator~(GLO), and a global estimator with simplicial
constraints~(GLOC). Furthermore, we demonstrate the robustness of our methods by
evaluating their performance under noisy conditions, showing that GLOC
achieves relatively higher accuracy, even with smaller observation data sets. By
capturing the evolving nature of interactions, our framework provides a systematic
and practical approach to reconstructing higher-order structures from dynamic
real-world data.

The remainder of the article is organized as follows:
in Section~\ref{Sec:Formalism}, we describe the formalism
of evolutionary games we adopt; in Section~\ref{Sec:Methods}
we rigorously formulate the reconstruction problem and
discuss~PBP, GLO and~GLOC in detail; in Section~\ref{Sec:Creation}
we briefly introduce the different methods we use to
create synthetic complexes, or to construct complexes
from empirical data, as well as the metrics we use to
assess the performance of our methods; in Section~\ref{Sec:Results}
we present the results of the three methods on synthetic
complexes and on empirical ones; finally, we give our
concluding remarks in Section~\ref{Sec:Discussion}.

\section{Formalism of evolutionary games}\label{Sec:Formalism}
Evolutionary dynamics is often used as a paradigmatic
model to study the formation of complex social interactions
between agents on a network. In general, the dynamics
of an evolutionary game advances via two steps, namely
computing the payoffs that players receive according
to the combination of their choices, and updating each
player's strategy according to the state of the system
at any given moment in time.

Here, we consider four classic games, namely the Prisoner's
Dilemma, the Stag Hunt, the Snowdrift and the Harmony game.
To reduce the computational costs, and focus on the effectiveness
of our methods, we only consider pairwise and three-body interactions.
However, our approaches can be straightforwardly generalized
to edges of any size. Also note that, conventionally, an edge
linking $d$~nodes is called a $d-1$-simplex. Thus, simple edges,
corresponding to pairwise interactions, are 1-simplices, whereas
edges representing triadic interactions are 2-simplices. All
four games we use represent situations in which players choose
between cooperative behaviour~(C) and individualistic defection~(D).
Thus, when a game is being played by 2~players at a time, i.e.,
on a 1-simplex, there are 4~possible cases, whereas the total
number of possible outcomes is~8 when the game is played by
3~players at a time, i.e., on a 2-simplex. As illustrated schematically
in Fig.~\ref{Fig:GraphicIllustration}, in pairwise games both
players receive a unitary reward when they cooperate with each
other, and receive nothing if they both defect. If one player
cooperates while the other defects, then the cooperating one
receives a so-called sucker's payoff~$S$, and the defecting
one receives the temptation payoff~$T$. In three-body games,
the players still receive a unitary reward or nothing for cooperating
or defecting all together, respectively, and a player's payoff
is still~$S$ or~$T$ when they are the only one to cooperate
or defect, respectively. However, in these last two cases,
the other two players receive a payoff~$W$ for defecting or
a payoff~$G$ for cooperating.

Note that the difference in the equilibrium states
of the four games is determined by the values of~$T$
and~$S$. Specifically, in the Prisoner's Dilemma,
for which $T>1$ and $S<0$, the only Nash equilibrium
is mutual defection. In the Harmony game, which has
the opposite relations $T<1$ and $S>0$, mutual cooperation
prevails. In the Snowdrift game, where $T>1$ and $S>0$,
the equilibrium states feature one player cooperating
and the other one defecting. Finally, in the Stag
Hunt, with $T<1$ and $S<0$, the equilibrium states
have both players either cooperating or defecting
at the same time. The Nash equilibria for three-body
games, however, are more complex, and depend on the
deviation between~$W$ and~$G$~\cite{Civ24}.

In a simplicial complex of~$N$ players,
denoting the strategy of player~$i$ at
round~$t_m$ as~$s_i(t_m)$, the total payoff
for player~$i$ at that round, $\pi_i(t_m)$,
is obtained by summing the individual contributions
from each simplex it participates in:
\begin{equation}
 \begin{split}
  \pi_i(t_m) &= \sum_{j=1}^N A^{(1)}_{i,j} P^{(1)}_{i,j}(t_m)\\
  &\quad + \sum_{\substack{j_1, j_2=1 \\ j_1<j_2}}^N A^{(2)}_{i,j_1,j_2} P^{(2)}_{i,j_1,j_2}(t_m)\:.
 \end{split}
\end{equation}
In the equation above, $\mathbf{A^{(1)}}$
and~$\mathbf{A^{(2)}}$ are the adjacency
matrix of the 1-simplices and the adjacency
tensor of the 2-simplices, respectively,
and their elements are~1 if an edge exists
on the nodes corresponding to their indices,
and~0 otherwise. Also, $\mathbf{P^{(1)}}(t_m)$
and~$\mathbf{P^{(2)}}(t_m)$ are the payoff
matrix and the payoff tensor for pairwise
and three-body interactions at time~$t_m$,
respectively, and their~$(i,j)$ or $(i,j_1,j_2)$~elements
are the reward that player~$i$ receives if
it participates in a game with player~$j$
or with players~$j_1$ and~$j_2$, given a
choice of strategies.

At each time step, players update their strategies based
on a simple imitation rule: random selection among neighbors.
Specifically, at the beginning of each round, each player
randomly selects one of their neighbors and adopts their
last strategy for the next turn. This rule ensures that
strategies do not rapidly converge to equilibrium, allowing
us to capture richer transient-phase dynamics over a longer
observation period~\cite{Han20}. In real-world social systems,
where equilibrium is often impractical due to the dynamic
and fluctuating nature of interactions, relying on
transient-phase data provides a more realistic representation
of higher-order structures.


\section{Simplicial complex reconstruction}\label{Sec:Methods}
\subsection{Problem formulation}
The goal of our higher-order network reconstruction
methods is to determine the structure of unknown simplicial
complexes from observations of the payoffs and of the
strategies of the nodes at~$M$ different times. In
other words, we aim to find~$\mathbf{A}^{(1)}$ and~$\mathbf{A}^{(2)}$
from the payoff matrix~$\boldsymbol\Pi$ and the strategy
matrix~$\mathbf S$, whose columns contain the time
series of payoffs and strategies for the corresponding
players, and have the form
\begin{equation}
    \boldsymbol\Pi =
    \begin{pmatrix}
    \pi_1(t_1) & \pi_2(t_1) & \cdots & \pi_N(t_1) \\
    \pi_1(t_2) & \pi_2(t_2) & \cdots & \pi_N(t_2) \\
    \vdots     & \vdots     & \ddots & \vdots \\
    \pi_1(t_M) & \pi_2(t_M) & \cdots & \pi_N(t_M) \\
    \end{pmatrix}\:,
\end{equation}
and
\begin{equation}
    \mathbf S =
    \begin{pmatrix}
    s_1(t_1) & s_2(t_1) & \cdots & s_N(t_1) \\
    s_1(t_2) & s_2(t_2) & \cdots & s_N(t_2) \\
    \vdots   & \vdots   & \ddots & \vdots \\
    s_1(t_M) & s_2(t_M) & \cdots & s_N(t_M) \\
    \end{pmatrix}\:.
\end{equation}

The first step in reaching this goal is to note that,
given~$M$ observations, all the equations concerning
player~$i$ can be rewritten collectively as a system
of linear equations. To do so, we introduce some auxiliary
quantities. First, we call~$\Pi_i$ and~$\hat\Pi_{t_m}$
the $i$-th column and the $t_m$-th row of $\boldsymbol\Pi$,
respectively:
\begin{align}
 \Pi_i         &= \begin{pmatrix} \pi_i(t_1), \pi_i(t_2), \dotsc, \pi_i(t_M)\end{pmatrix}^\mathrm{T}\:,\\
 \hat\Pi_{t_m} &= \begin{pmatrix} \pi_1(t_m), \pi_2(t_m), \dotsc, \pi_N(t_m)\end{pmatrix}\:.
\end{align}
Then, we build the matrices of time series of pairwise
and three-body payoffs for player~$i$,
\begin{equation}
 \boldsymbol\Phi_i^{(1)} = \begin{pmatrix}
                            P_{i,1}^{(1)}(t_1) & P_{i,2}^{(1)}(t_1) & \cdots & P_{i,N}^{(1)}(t_1) \\
                            P_{i,1}^{(1)}(t_2) & P_{i,2}^{(1)}(t_2) & \cdots & P_{i,N}^{(1)}(t_2) \\
                            \vdots             & \vdots             & \ddots & \vdots \\
                            P_{i,1}^{(1)}(t_M) & P_{i,2}^{(1)}(t_M) & \cdots & P_{i,N}^{(1)}(t_M)
                           \end{pmatrix}
\end{equation}
and
\begin{widetext}
\begin{equation}
 \boldsymbol\Phi_i^{(2)} = \begin{pmatrix}
                            P_{i,1,1}^{(2)}(t_1) & P_{i,1,2}^{(2)}(t_1) & \cdots & P_{i,1,N}^{(2)}(t_1) & P_{i,2,1}^{(2)}(t_1) & \cdots & P_{i,N,N}^{(2)}(t_1)\\
                            P_{i,1,1}^{(2)}(t_2) & P_{i,1,2}^{(2)}(t_2) & \cdots & P_{i,1,N}^{(2)}(t_2) & P_{i,2,1}^{(2)}(t_2) & \cdots & P_{i,N,N}^{(2)}(t_1)\\
                            \vdots               & \vdots               & \ddots & \vdots               & \vdots               & \ddots & \vdots \\
                            P_{i,1,1}^{(2)}(t_M) & P_{i,1,2}^{(2)}(t_M) & \cdots & P_{i,1,N}^{(2)}(t_2) & P_{i,2,1}^{(2)}(t_M) & \cdots & P_{i,N,N}^{(2)}(t_M)\\
                           \end{pmatrix}\:.
\end{equation}
\end{widetext}
Last, we build the adjacency flattenings for player~$i$:
\begin{equation}
 a_i^{(1)} = \begin{pmatrix} A_{i,1}^{(1)}, A_{i,2}^{(1)}, \dotsc, A_{i,N}^{(1)}\end{pmatrix}^\mathrm{T}
\end{equation}
and
\begin{equation}
 a_i^{(2)} = \begin{pmatrix} A_{i,1,1}^{(2)}, A_{i,1,2}^{(2)}, \dotsc, A_{i,1,N}^{(2)}, A_{i,2,1}^{(2)}, \dotsc, A_{i,N,N}^{(2)},\end{pmatrix}^\mathrm{T}\:.
\end{equation}

Then, introducing the notations $\mathbf{\Phi}_i = \begin{pmatrix} \mathbf{\Phi}_i^{(1)}, \mathbf{\Phi}_i^{(2)}\end{pmatrix}$
and $A_i = \begin{pmatrix}\lp a_i^{(1)}\rp^\mathrm{T}, \lp a_i^{(2)}\rp^\mathrm{T}\end{pmatrix}^\mathrm{T}$, we can write
\begin{equation}\label{Eq:LinearEquations}
    \Pi_i = \boldsymbol\Phi_i \cdot A_i.
\end{equation}
Note that, even though the strategies do not feature
explicitly in the equation above, the elements of~$\boldsymbol\Phi_i$
and those of~$\Pi_i$ do depend on the choices of all
the players that are connected to player~$i$. Thus,
the strategy matrix is implicitly entangled with the
payoff one.

With this setup, solving the system in Eq.~\ref{Eq:LinearEquations}
for $A_i$ allows one to reconstruct all the 1-simplices and 2-simplices
in which player~$i$ participates, and the entire topology of the network
can be estimated by repeating the procedure for every player. Then,
a fundamental requirement is to ensure that the results obtained from
the iterations of the procedure are consistent amongst themselves and
with the nature of the network being reconstructed. To accomplish this,
we propose three methods. First, we build a point-by-point estimator~(PBP)
to locally infer $A_i$; then, we extend it to a global estimator~(GLO);
finally, we add simplicial constraints~(GLOC), increasing the reconstruction
accuracy for noisy data.

\subsection{Point-by-point estimator}
In the context of network reconstruction, the point-by-point
(PBP) estimator is a local approach that infers the structure of a network by
independently solving linear equations for each node~\cite{Wan11,Su12,Wan16, Han15}, based on observed
dynamical data~\cite{Yeu02,Xia18}. This method involves estimating the connections of each
node separately~\cite{Tim07,She14,Li17}, without considering the global structure of the network~\cite{Che18,Ma23}.
Specifically, assuming that the unknown network is sparse,
a solution to Eq.~\ref{Eq:LinearEquations} is
given by
\begin{equation}
 \mathop{\arg\min}_{\lbr A_i\rbr\ |\ \Pi_i = \boldsymbol\Phi_i \cdot A_i}\lp\left\| A_i\right\|_1\rp\:,
\end{equation}
where $\left\|A_i\right\|_1$ is the 1-norm
of~$A_i$, i.e., the sum of its absolute values.
Reconstruction of the whole network results
then from the union of the edges detected
by solving the system for each individual
node. Thus, this method is a point-by-point
estimator~(PBP) that builds a network from
the local neighbourhoods of all the nodes.
To mitigate the effect of noise on the measurements
of~$\Pi_i$, we include a 2-norm penalty in
the equation above~\cite{Ma23}, so that the
solution we seek is
\begin{equation}\label{Eq:PBP_L1&2_Penalty}
 \mathop{\arg\min}_{\lbr A_i\rbr\ |\ \Pi_i = \boldsymbol\Phi_i \cdot A_i}\lp\left\| A_i\right\|_1 + \left\| \Pi_i - \boldsymbol\Phi_i\cdot A_i\right\|_2\rp\:.
\end{equation}
Note that in this last equation the 1-norm
is a structural quantity, as the elements
of~$A_i$ constitute the local topology of
the network, whereas the 2-norm is a temporal
one, because the elements of~$\Pi_i$ relate
to measurements at different moments.

We solve the last equation by applying
the Orthogonal Matching Pursuit method~\cite{Hua20}.
Then, one would typically assign an edge
between node~$i$ and node~$j$, or one
between~$i$, $j$ and~$k$, if the corresponding
element of~$a_i^{(1)}$ or~$a_i^{(2)}$
is greater than~0. However, the presence of
higher-order interactions introduces significant
sensitivity to noise~\cite{Wan22}, even when a 2-norm penalty
is employed, as formulated in Eq.~\ref{Eq:PBP_L1&2_Penalty}.
To mitigate this issue, we define adaptive thresholds~$\Delta_i^{(1)}$
for simple edges and~$\Delta_i^{(2)}$ for 2-simplices,
retaining only elements of~$a_i^{(1)}$ and~$a_i^{(2)}$ that
exceed these thresholds. The thresholds are computed using
a gap-based approach~\cite{Ma18,2Ma23}. Specifically, the elements of~$a_i^{(1)}$
and~$a_i^{(2)}$ are sorted in non-increasing order. Next, the gap between consecutive elements in each sorted vector is measured, weighted by the ratio of the same consecutive elements. Finally, the threshold values are determined by selecting the largest weighted gap, ensuring robustness against noise while preserving essential structural information~\cite{Li17}.
In formulae,
\begin{equation}\label{Eq:Delta}
 \Delta_i^{(d)} = \mathop{\arg\max}_{h} \lp \frac{{a'}_{i,h}^{(d)}}{{a'}_{i,h+1}^{(d)}} \lp{a'}_{i,h}^{(d)} - {a'}_{i,h+1}^{(d)} \rp\rp\:,
\end{equation}
where~$d$ is the dimension of the simplices
(1 or~2), and~${a'}_i^{(1)}$ and~${a'}_i^{(2)}$
are the sorted versions of~$a_i^{(1)}$ and~$a_i^{(2)}$.
Finally, to prevent inconsistencies in the
reconstructed simplicial complex, we drop any
2-simplex $(i, j, k)$ if any edge $(i, j)$,
$(i, k)$ or $(j, k)$ is missing from the network.

\subsection{Global estimator}
Even when eliminating simplicial inconsistencies,
the PBP can still lead to reconstruction errors,
because of the indepencence of the estimates of different
nodes. Thus, for example, the edge~$(i,j)$, corresponding
to the element~$A_{i,j}^{(1)}$, may end up being
included in the network, while the edge~$(j,i)$,
corresponding to the element~$A_{j,i}^{(1)}$, is
excluded, which would be inconsistent with the fact
that the simplicial complex being reconstructed is
undirected, and, as such, it must satisfy
\begin{equation}\label{Eq:SymmetricConstraint_1}
 A_{i,j}^{(1)}=A_{j,i}^{(1)}\:.
\end{equation}
The situation quickly becomes more complex with
2-simplices, where the constraints to be satisfied
are
\begin{equation}\label{Eq:SymmetricConstraint_2}
 A_{i,j,k}^{(2)} = A_{i,k,j}^{(2)} = A_{j,i,k}^{(2)} = A_{j,k,i}^{(2)} = A_{k,i,j}^{(2)} = A_{k,j,i}^{(2)}\:.
\end{equation}

To address the limitations of~PBP, the global estimator (GLO)
reconstructs the network by solving all equations simultaneously
while enforcing structural constraints~\cite{Han20}. Unlike local approaches
that estimate each node independently, the global estimator
treats the entire network as a coupled system~\cite{Gu22,Mal24}. This method
formulates network reconstruction as a large-scale optimization
problem~\cite{Liu23}, where the adjacency matrix is inferred by minimizing
discrepancies between observed dynamical data and predicted interactions~\cite{Han15,Bru16,Shi21}.
To do so, we first introduce compressed flattenings
of~$\mathbf{A^{(1)}}$ and~$\mathbf{A^{(2)}}$, which
exclude any element on any diagonal and include only
the upper triangle and the uppermost pyramid, respectively,
so that
\begin{equation}
 \hat A^{(1)} = \begin{pmatrix} A_{1,2}^{(1)}, A_{1,3}^{(1)}, \dotsc, A_{1,N}^{(1)}, A_{2,3}^{(1)}, \dotsc, A_{N-1,N}^{(1)}\end{pmatrix}^\mathrm{T}
\end{equation}
and
\begin{equation}
 \hat A^{(2)} = \begin{pmatrix} A_{1,2,3}^{(2)}, A_{1,2,4}^{(2)}, \dotsc, A_{1,2,N}^{(2)}, A_{1,3,4}^{(2)}, \dotsc, A_{N-2,N-1,N}^{(2)}\end{pmatrix}^\mathrm{T}\:.
\end{equation}
Then, we use these to build the vector of unknowns~$\hat{A}$,
\begin{equation}\label{Eq:A_hat}
    \hat A = \begin{pmatrix} \lp\hat A^{(1)}\rp^\mathrm{T}, \lp\hat A^{(2)}\rp^\mathrm{T}\end{pmatrix}^\mathrm{T}\:,
\end{equation}
which implicitly contains all the needed symmetry constraints.

Next, we carry out a similar procedure
on the~$\Pi_i$ and the~$\boldsymbol\Phi_i$.
This results in the vectors
\begin{equation}
 \begin{split}
  \hat\Pi &= \begin{pmatrix} \hat\Pi_{t_1}, \hat\Pi_{t_2}, \dotsc, \hat\Pi_{t_M}\end{pmatrix}^\mathrm{T} \\
  &= \lp \pi_1(t_1), \pi_2(t_1), \cdots, \pi_N(t_1), \pi_1(t_2), \cdots,\pi_N(t_M)\rp^\mathrm{T}
 \end{split}
\end{equation}
and
\begin{equation}\label{Eq:Phi_hat}
 \boldsymbol{\hat\Phi} = \begin{pmatrix} \boldsymbol{\hat\Phi}_{t_1}, \boldsymbol{\hat\Phi}_{t_2}, \dotsc, \boldsymbol{\hat\Phi}_{t_M}\end{pmatrix}^\mathrm{T}\:.
\end{equation}
In the equation below,
\begin{widetext}
 \begin{equation}
 \boldsymbol{\hat\Phi}_{t_m} =
    \begin{pmatrix}
     C^{(1)}_{1,2,1}(t_m)       & C^{(1)}_{1,2,2}(t_m)       & \cdots & C^{(1)}_{1,2,N}(t_m) \\
     C^{(1)}_{1,3,1}(t_m)       & C^{(1)}_{1,3,2}(t_m)       & \cdots & C^{(1)}_{1,3,N}(t_m) \\
     \vdots                     & \vdots                     & \ddots & \vdots \\
     C^{(1)}_{1,N,1}(t_m)       & C^{(1)}_{1,N,2}(t_m)       & \cdots & C^{(1)}_{1,N,N}(t_m) \\
     C^{(1)}_{2,3,1}(t_m)       & C^{(1)}_{2,3,2}(t_m)       & \cdots & C^{(1)}_{2,3,N}(t_m) \\
     \vdots                     & \vdots                     & \ddots & \vdots \\
     C^{(1)}_{N-1,N,1}(t_m)     & C^{(1)}_{N-1,N,2}(t_m)     & \cdots & C^{(1)}_{N-1,N,N}(t_m) \\
     C^{(2)}_{1,2,3,1}(t_m)     & C^{(2)}_{1,2,3,2}(t_m)     & \cdots & C^{(2)}_{1,2,3,N}(t_m) \\
     C^{(2)}_{1,2,4,1}(t_m)     & C^{(2)}_{1,2,4,2}(t_m)     & \cdots & C^{(2)}_{1,2,4,N}(t_m) \\
     \vdots                     & \vdots                     & \ddots & \vdots \\
     C^{(2)}_{1,2,N,1}(t_m)     & C^{(2)}_{1,2,N,2}(t_m)     & \cdots & C^{(2)}_{1,2,N,N}(t_m) \\
     C^{(2)}_{1,3,4,1}(t_m)     & C^{(2)}_{1,3,4,2}(t_m)     & \cdots & C^{(2)}_{1,3,4,N}(t_m) \\
     \vdots                     & \vdots                     & \ddots & \vdots \\
     C^{(2)}_{N-2,N-1,N,1}(t_m) & C^{(2)}_{N-2,N-1,N,2}(t_m) & \cdots & C^{(2)}_{N-2,N-1,N,N}(t_m) \\
    \end{pmatrix}\:,
\end{equation}
\end{widetext}

\clearpage
where
\begin{equation}
 C^{(1)}_{i,j,n}(t_m) = \begin{cases}
                         P^{(1)}_{i,j}(t_m) &\quad\text{if $n=i$}\\[0.2cm]
                         P^{(1)}_{j,i}(t_m) &\quad\text{if $n=j$}\\[0.2cm]
                         0 &\quad\text{otherwise}
                        \end{cases}
\end{equation}
and
\begin{equation}
 C^{(2)}_{i,j,k,n}(t_m) = \begin{cases}
                           P^{(2)}_{i,j,k}(t_m) &\quad\text{if $n=i$}\\[0.2cm]
                           P^{(2)}_{j,i,k}(t_m) &\quad\text{if $n=j$}\\[0.2cm]
                           P^{(2)}_{k,i,j}(t_m) &\quad\text{if $n=k$}\\[0.2cm]
                           0 &\quad\text{otherwise}\:.
                          \end{cases}
\end{equation}

With this formalism, it is now possible
to conduct a one-shot optimization of the
entire network structure, aiming to find
the solution given by
\begin{equation}\label{Eq:GLO_L1&2_Penalty}
 \mathop{\arg\min}_{\lbr \hat A\rbr\ |\ \hat\Pi = \boldsymbol{\hat\Phi} \cdot \hat A}\lp\left\| \hat A\right\|_1 + \left\| \hat\Pi - \boldsymbol{\hat\Phi}\cdot\hat A\right\|_2\rp\:.
\end{equation}
Thus, this method represents a global estimator~(GLO).

After estimating~$\hat A$, we then sort~$\hat A^{(1)}$
and~$\hat A^{(2)}$ and determine the two global thresholds~$\hat\Delta^{(1)}$
and~$\hat\Delta^{(2)}$, using the same function as in
Eq.~\ref{Eq:Delta}, and delete all reconstructed edges
whose corresponding element is smaller than or equal to
the appropriate threshold. Finally, we delete any 2-simplex
that has at least one missing 1-simplex on its constituent
nodes, to guarantee that the reconstructed simplicial
complex is consistent.

\subsection{Global estimator with simplicial constraints}
Our last method extends the global estimator to include
an additional set of simplicial constraints~(GLOC). Specifically,
we maintain the formalism of~GLO, but add the requirement
that the value of the element of~$\hat A^{(2)}$ corresponding
to a 2-simplex be bounded from above individually by the
three values of the elements of~$\hat A^{(1)}$ corresponding
to the 1-simplices on the same nodes, and that all be less
than or equal to~1. This guarantees that a 2-simplex that
is highly likely to exist implies 1-simplices on the same
nodes that are at least as likely to be there. In fact,
in the limiting case that an edge~$(i,j,k)$ exists for
sure, all three edges~$(i,j)$, $(i,k)$ and $(j,k)$ will
certainly exist as well. The solution to find can then
be written as
\begin{equation}\label{Eq:GLOC_L1&2_Penalty}
 \mathop{\arg\min}_{\substack{\lbr \hat A\rbr\ |\ \hat\Pi = \boldsymbol{\hat\Phi} \cdot \hat A\\ A_{i,j,k}^{(2)}\leqslant A_{i,j}^{(1)}\leqslant 1\\ A_{i,j,k}^{(2)}\leqslant A_{i,k}^{(1)}\leqslant 1\\ A_{i,j,k}^{(2)}\leqslant A_{j,k}^{(1)}\leqslant 1}}\lp\left\| \hat A\right\|_1 + \left\| \hat\Pi - \boldsymbol{\hat\Phi}\cdot\hat A\right\|_2\rp\:.
\end{equation}

After solving the problem numerically, we still carry
out the thresholding procedure as described above. As
this can introduce structural inconsistencies, we also
check for 2-simplices with missing induced 1-simplices
and delete any we find, as in the previous methods.

\section{Network construction and performance metrics}\label{Sec:Creation}
To validate our methods, we conduct numerical simulations
on three different types of synthetic complexes and on three
complexes extracted from empirical data. Note that the structure
of these complexes is used to advance the evolutionary games,
but no information about it takes part in the reconstruction
process.

\subsection{Synthetic complexes}\label{Sec:SynSC}
\subsubsection{Non-preferential attachment model (NPA)}
To build a simplicial complex in which the construction
process does not include any bias, we follow the procedure
outlined in Ref.~\cite{Kov21}. Starting from an initial
kernel of $N_0=5$~fully-connected nodes, we grow the complex
by adding one node at each time step~$t$. When a new node
is added, $m_\mathrm{tri}$ existing edges are randomly
extracted, and the new node is linked to the nodes that
form them. Note that if $m_\mathrm{tri}>1$, we choose the
links without any pairwise adjacency, so that the total
number of nodes to which the new one is to be linked is
always~$2m_\mathrm{tri}$. Following this procedure, at
time~$t$ each edge has the same probability $p=\ls\frac{N_0\lp N_0 - 1\rp}{2} + 2 m_{\text{tri}}\lp t - 1\rp\rs^{-1}$
to be extracted. Thus, no preferential attachment takes
place in this model. After reaching the desired number
of nodes~$N$, we take a fraction~$\rho$ of all triangles
and add 2-simplices on the nodes that form them.

\subsubsection{Preferential attachment model (PA)}
To create complexes in which the growth process
follows a preferential attachment rule, we change
the probability distribution of the edges selected
for linking the new node from uniform to one that
is proportional to the number of triangles that
each link is part of. Thus, the probability of
extracting the edge $(i,j)$ is $p_{ij}=\frac{k_{ij}}{\sum_{(i,j)}k_{ij}}$,
where~$k_{ij}$ is the number of triangles that
edge $(i,j)$ participates in, and the sum is on
all edges currently in the network. Note that the
quantities in the previous formula depend on the
time step, and we have avoided explicitly writing
this dependence to reduce clutter. Also in this
case, upon reaching the target size, we extract
a random fraction~$\rho$ of triangles and add 2-simplices
on their nodes.

\subsubsection{Mixed model (MIX)}
The two models above can be further integrated
and generalized into a mixed one, which allows
one to tune the power-law exponents of the resulting
degree distributions for simple edges and 2-simplices~\cite{Kov21}.
To achieve this, we turn the probability of selecting
the edge $(i,j)$ into $p_{ij} = \lp 1-\frac{3}{2}B\rp\ls\frac{N_0\lp N_0 - 1\rp}{2} + 2 m_{\text{tri}}\lp t - 1\rp\rs^{-1} + B\frac{k_{ij}}{\sum_{(i,j)}k_{ij}}$. Here, the coefficient~$B$ is allowed
to vary between~0 and~2, and it is worth noticing
that the model reduces to NPA and PA for $B=0$
and $B=\frac{2}{3}$, respectively. As in the previous
two models, once the network has~$N$ nodes, we
add 2-simplices on the nodes that form a fraction~$\rho$
of all the triangles.

\subsection{Empirical complexes}\label{Sec:Empirical}
To demonstrate the applicability of our reconstruction
methods, we also use simplicial complexes inferred from
real-world data. Specifically, we consider the network
of social contacts recorded in a rural village in Malawi
(Malawi)~\cite{Oze21}, the high-resolution data of face-to-face
contacts during a scientific conference (SFHH)~\cite{Gen18}
and the interactions between patients and healthcare workers
over a period of 96~hours in a non-emergency unit of a
hospital in Lyon (LH10)~\cite{Van13}. To build a simplicial
complex from each of the data sets, we first generate
a weighted network in which the weights correspond to
the total number of observed interactions between node
pairs. Then, we delete all edges with a weight smaller
than or equal to a threshold~$\zeta$ and restrict ourselves
to the unweighted version of the resulting largest connected
component. The values we use for the threshold are~35
for Malawi, 55 for SFHH and~60 for LH10, resulting in
networks of~50, 54 and~49 nodes, respectively. Finally,
we split the data into 5-minute time windows. Any three
nodes that form a triangle in any temporal window are
turned into a 2-simplex.

\subsection{Reconstruction and performance metrics}
To evaluate the performance of our methods
in reconstructing the networks, we use several
metrics, based on the number of edges that
are correctly or incorrectly detected. To
define them, we start by calling~TP the number
of true positives, i.e., the number of existing
edges that are determined. We also indicate
with~FP the number of false positives, which
are edges that are incorrectly identified
to exist. We use~TN for the number of true
negatives, which are missing edges that are
correctly detected not to exist. Finally,
FN is the number of false negatives, which
are edges that are present in the network,
but that a method fails to detect. Note that,
the sum of these four values is the total
possible number of edges~$E$:
\begin{multline}
 \TP+\TN+\FP+\FN = E\\ = \begin{cases}
                        \binom{N}{2} = \frac{N\lp N-1\rp}{2} &\quad\text{for 1-simplices}\\[0.2cm]
                        \binom{N}{3} = \frac{N\lp N-1\rp\lp N-2\rp}{6} &\quad\text{for 2-simplices.}
                       \end{cases}
\end{multline}
Then, we evaluate the results using accuracy~(ACC),
true positive rate~(TPR), positive predictive
value~(PPV), true negative rate~(TNR) and~$F_1$
score, which are defined as follows:
\begin{align}
 \mathrm{ACC} &= \frac{\TP + \TN}{E}\:,\\
 \mathrm{TPR} &= \frac{\TP}{\TP + \FN}\:,\\
 \mathrm{PPV} &= \frac{\TP}{\TP + \FP}\:,\\
 \mathrm{TNR} &= \frac{\TN}{\TN + \FP}\:,\\
 F_1 &= \frac{2\TP}{2\TP + \FP + \FN}\:.
\end{align}
The range of all these quantities is between~0
and~1, with larger values indicating a better
performance. Additionally, we consider the Matthews
correlation coefficient~(MCC), defined as
\begin{equation}
 \mathrm{MCC} = \frac{\TP\cdot\TN - \FP\cdot\FN}{\sqrt{\lp\TN + \FN\rp\lp\TN + \FP\rp\lp\TP + \FP\rp\lp\TP + \FN\rp}},
\end{equation}
whose range is $[-1, 1]$.

\begin{figure*}[t]
\centering
\includegraphics[width=0.95\textwidth]{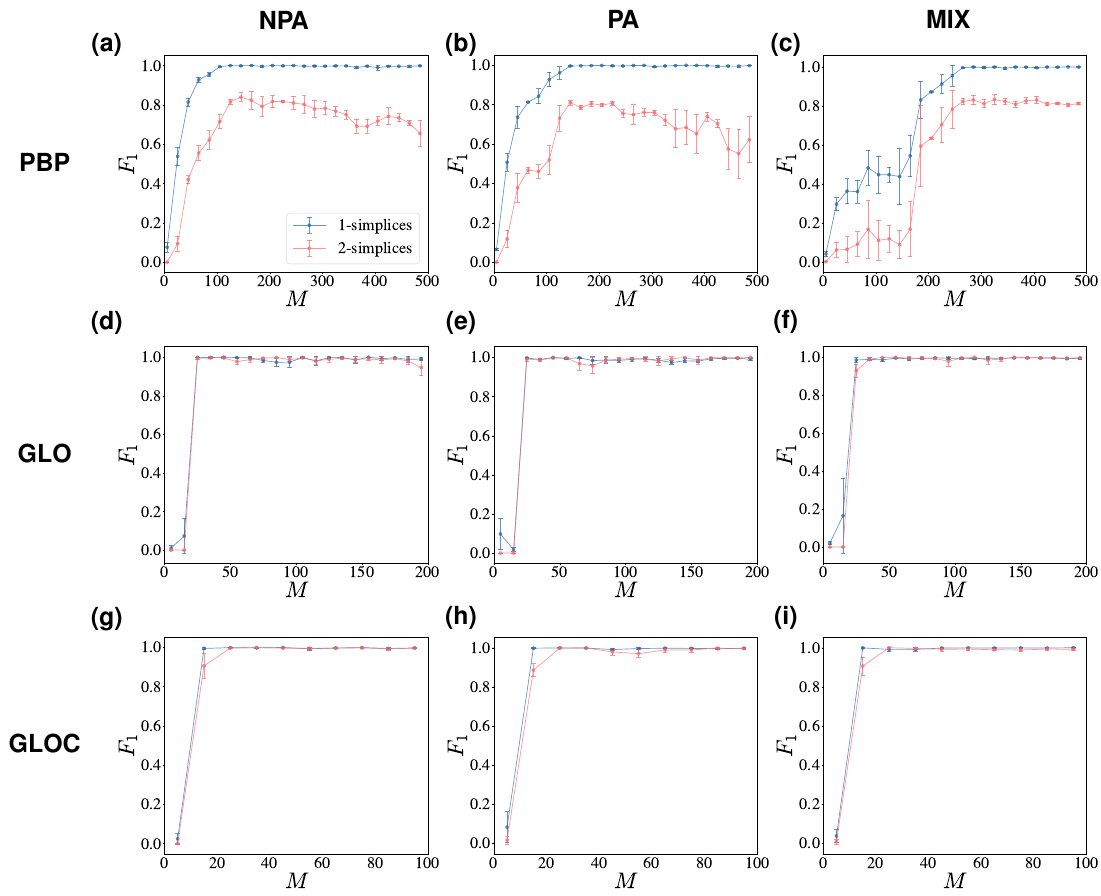}
\caption{\textbf{Global reconstruction methods outperform PBP.}
All three methods reach the highest possible value of the $F_1$~score
for the reconstruction of 1-simplices (blue) playing the Prisoner's
Dilemma game for large enough sizes~$M$ of the observation sets.
However, the score of PBP for 2-simplices (red) never increases
above~$0.8$. At the same time, GLO and GLOC require fewer than
30 observations to reach a score of~1.}\label{Fig:Syn_F1}
\end{figure*}
Furthermore, using the false positive rate
\begin{equation}
 \mathrm{FPR} = \frac{\FP}{\FP+\TN}\:,
\end{equation}
the recall rate
\begin{equation}
 \mathrm R = \frac{\TP}{\TP+\FN}\:,
\end{equation}
and the precision rate
\begin{equation}
 \mathrm P = \frac{\TP}{\TP+\FP}\:,
\end{equation}
we can define the receiver operating characteristic
and the precision-recall curve. The former
expresses how the TPR changes as a function
of the FPR, and the latter how precision decreases
as recall increases. Typically, the area under
these curves is computed, and used as a
measure of the classification performance
of a method, with values closer to the theoretical
maximum of~1 signifying better performance.
Note that, in our case, the results of the
method do not depend on a variable parameter.
Thus, the areas under these curves can be
computed analytically, so that the one under
their ROC curve is
\begin{equation}
 \mathrm{AUROC} = \frac{1}{2}\lp\frac{\TN}{\FP+\TN}+\frac{\TP}{\TP+\FN}\rp
\end{equation}
and the one under the precision-recall curve is
\begin{equation}
 \mathrm{AUPR} = \frac{1}{2}\frac{\TP\lp 2\TP + \FP + \FN\rp}{\lp\TP + \FP\rp\lp\TP + \FN\rp}\:.
\end{equation}

\section{Numerical results}\label{Sec:Results}
\subsection{Synthetic complexes}
\begin{figure*}[t]
\centering
\includegraphics[width=0.95\textwidth]{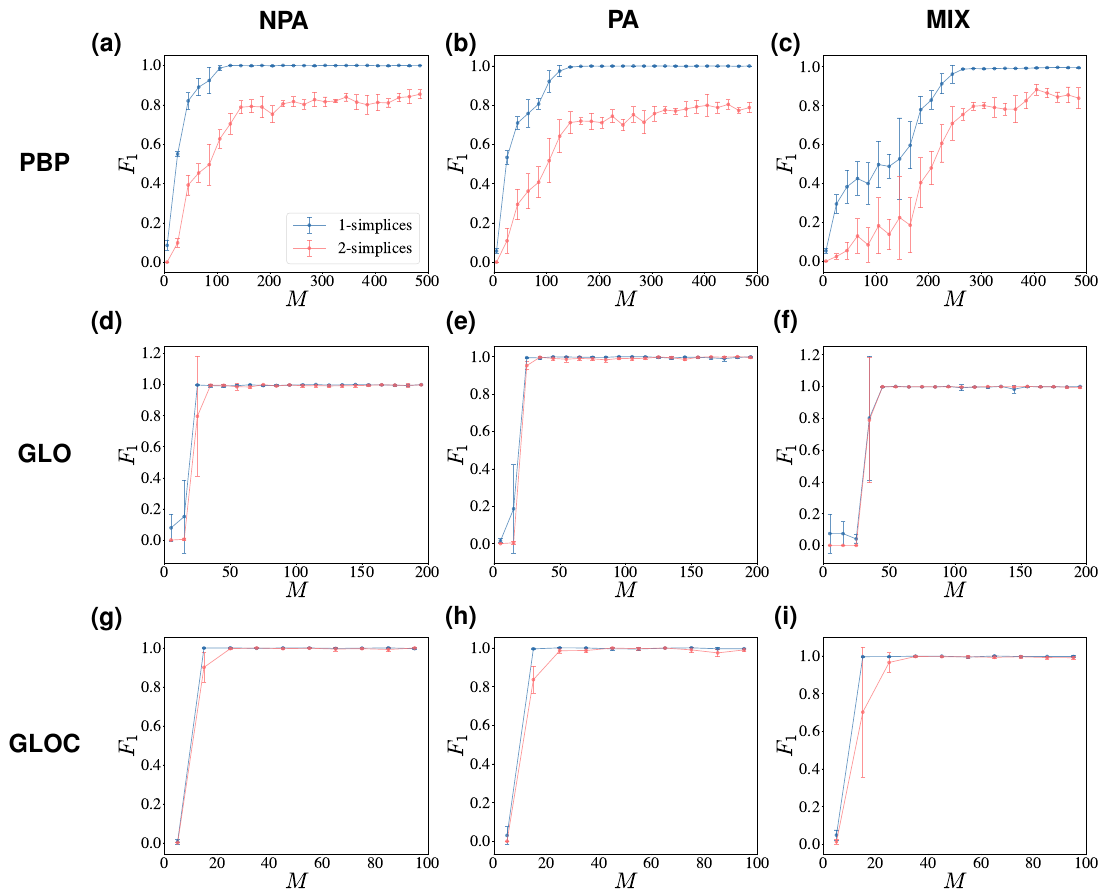}
\caption{\textbf{GLOC requires the smallest minimum observation set for noisy data.}
The relative performance of all three methods on noisy data ($q=1\%$) from a Prisoner's
Dilemma game is unchanged with respect to the noiseless case (Fig.~\ref{Fig:Syn_F1}).
However, the minimum size of the observation set~$M$ needed to achieve a high $F_1$~score
is higher than the one needed in the absence of noise, particularly for 2-simplices
(red) in the MIX model. Also, this increase is the smallest for GLOC, making it the
best of the three methods when data are affected by noise.}\label{Fig:Syn_F1_SIG1}
\end{figure*}
For our simulations, we use the synthetic
models described in Subsection~\ref{Sec:SynSC},
with $N = 50$~nodes and $\rho = 0.5$. For
the MIX model, we put $B=2$. In terms of
the evolutionary games, we set $T = 1.3$
and $S = -0.4$ for the Prisoner's Dilemma,
$T = 1.3$ and $S = 0.4$ for the Snowdrift
game, $T = 0.7$ and $S = -0.3$ for the Stag
Hunt game, and $T = 0.7$ and $S = 0.3$ for
the Harmony game. For the additional parameters
of the three-body games, we put $G = \frac{3}{5}\lp T + S\rp$
and $W = \frac{2}{5}\lp T + S\rp$.
Then, the payoff matrix and the strategy
matrix are recorded for 500~rounds during
five independent realization. Note that the
strategies do not converge to an equilibrium
within this number of rounds, because of
the update process we use.

\begin{figure*}[t]
\centering
\includegraphics[width=0.95\textwidth]{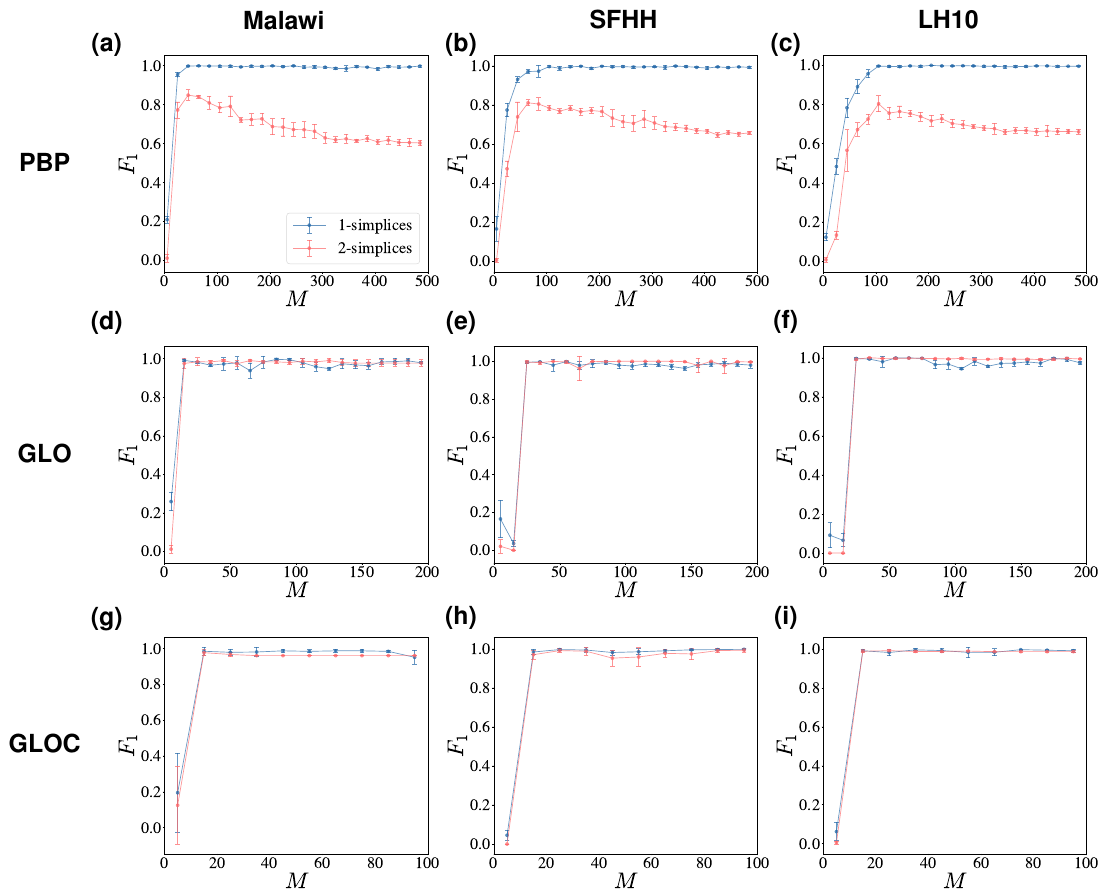}
\caption{\textbf{GLO and GLOC outperform PBP on empirical complexes.}
On empirical simplicial complexes, all three methods reach the highest possible value of the
$F_1$~score for the reconstruction of 1-simplices (blue) playing the Prisoner's Dilemma game
for large enough sizes~$M$ of the observation sets. However, the score of PBP (red) never increases
above~$0.85$ for 2-simplices.}\label{Fig:Emp_F1}
\end{figure*}
The reconstruction results for the synthetic
simplicial complexes are detailed in
Table~S1 of the Supplementary Material. For all
four games, all three methods provide very reliable
reconstructions, with the accuracy always very
close to~1. The two global estimators, however,
always perform better than PBP, especially when
comparing more sensitive metrics, such as the~MCC
and the $F_1$~score. The superiority of~GLO
and~GLOC over~PBP is even more evident in the
reconstruction of 2-simplices, for which the
differences in the complex metrics between the
methods are larger.

\begin{figure*}[t]
\centering
\includegraphics[width=0.95\textwidth]{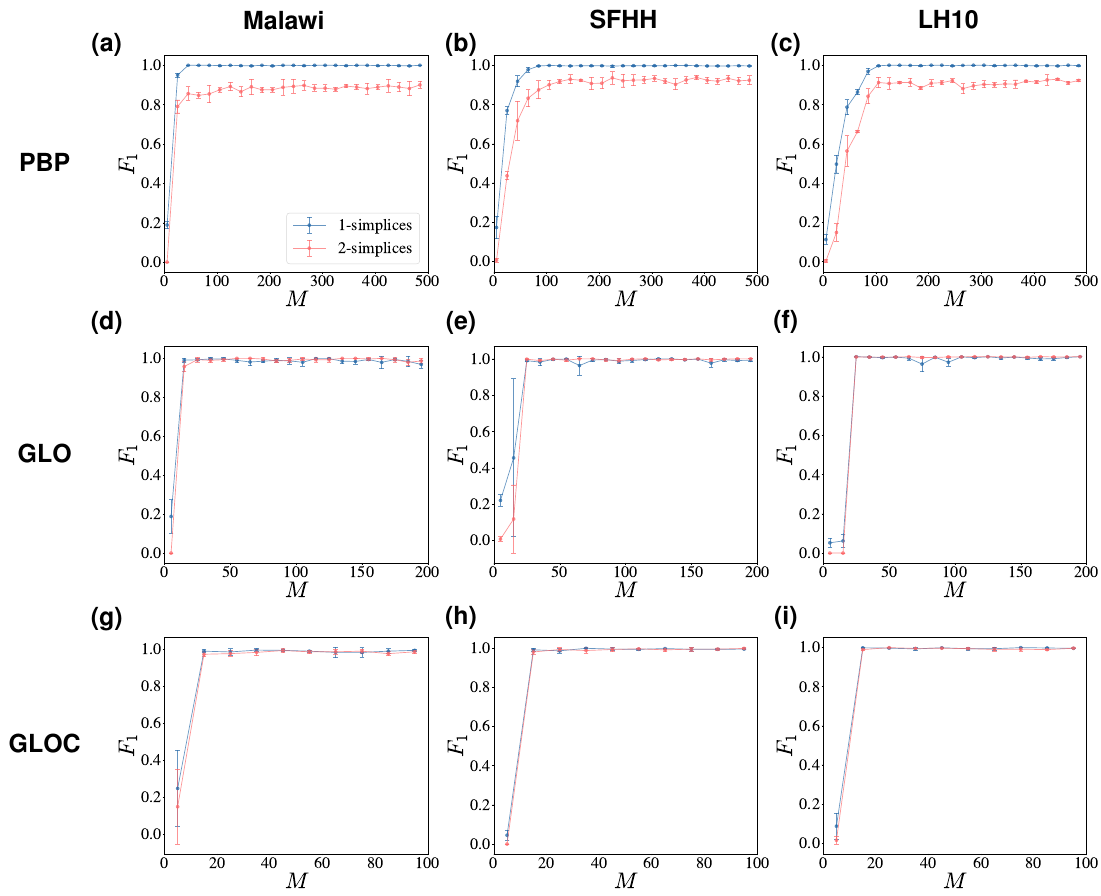}
\caption{\textbf{GLOC is the best performing method to
reconstruct empirical complexes with contaminated payoffs.}
On empirical simplicial complexes, the relative performance of all three methods on
noisy data ($q=1\%$) from a Prisoner's Dilemma game is the same as that of the noiseless
case (Fig.~\ref{Fig:Emp_F1}). However, GLO and GLOC require the smallest data sets
to achieve a high score. Moreover, the error bars for GLOC results are smaller than
those for GLO, indicating that it is more robust to noise.}\label{Fig:Emp_F1_SIG1}
\end{figure*}
To verify how the performance of the methods
changes with the size of the observation set,
we measured the $F_1$~score for different values
of~$M$. As shown in Fig.~\ref{Fig:Syn_F1} for
the Prisoner's Dilemma, all three methods reach
a high score for 1-simplices when~$M$ is large
enough. The difference between them lies in
their performance on 2-simplices, for which
the highest value that PBP achieves is never
above~$0.8$, and often with a large statistical
uncertainty. This indicates that the reconstructions
provided by PBP cannot be improved beyond a
certain quality by indefinitely increasing the
number of observations. Conversely, both GLO
and GLOC reach a score of~1 for all models with
fewer than 30~observations.

\subsection{Synthetic complexes with noise}
In real-world situations, empirical measurements in transient phases can be affected 
by different types of errors and uncertainties, leading to 
further challenges to reconstruct the topology of networks. 
To examine how perturbations affect the reliability of our methods, we tested them on "noisy" data using two different 
noise injection approaches. In one aspect, we created the noise of payoffs
by adding a random value~$\varepsilon$ to each
element of the measured payoff matrix~$\boldsymbol\Pi$.
Specifically, $\varepsilon$ was extracted from a
normal distribution with 0~mean and standard deviation
$\sigma=q\sigma_\Pi$, where $\sigma_\Pi$ is the
standard deviation of the distribution of elements
of the unperturbed payoff matrix. Additionally, we incorporated
flipping noise by inverting each element of the
strategy matrix~$\mathbf S$ with a probability $\psi$,
simulating errors in empirical observations
of strategies. This dual noise modeling allows us to
provide a more comprehensive assessment of robustness
beyond equilibrium states.

The reconstruction results for complexes with noise in payoffs,
detailed in Table~S2 of the Supplementary Material,
show that small amounts of noise ($q=1\%$) do not significantly
affect the accuracy of the methods, or their overall
performance. However, differences between PBP, GLO
and GLOC become more apparent when the amount of noise
is increased to $q=5\%$. In that case, and even for
1-simplices, only GLOC maintains high values across
the metrics. Indeed, GLOC consistently achieves the
highest scores for 2-simplices as well, demonstrating
its superiority in reconstructing higher-order interactions
regardless of the level of noise. Further analysis of strategy contamination noise,
as shown in Table~S3 of the Supplementary Material,
reveals that GLOC remains the most robust method,
even when the observed strategies are corrupted by
flipping errors. While all methods, especially GLO,
experience performance degradation due to noise in
strategies, GLOC still retains a significant advantage
over PBP and GLO.

The presence of noise also affects the minimum number
of required observations to achieve a high-quality reconstruction.
In fact, when noises in payoffs are emphasized (Fig.~\ref{Fig:Syn_F1_SIG1}),
the minimum size of the observation data set needed to obtain
an $F_1$~score of at least~$0.8$ for 2-simplices increases,
with respect to noiseless data, even if the noise factor
is only~$1\%$. 
When the noise originates from strategy contamination
(Fig.~S1 of the Supplementary Material),
the performance of GLO drops and fluctuates significantly across all three synthetic networks, with large error bars indicating high sensitivity to noisy strategies. PBP, while more stable, exhibits
a slight increase in $F_1$ scores with additional observations.
In contrast, GLOC achieves superior performance with far fewer observations, reaching an $F_1 \geq 0.4$ much earlier than
the other methods. These results
demonstrate the necessity of having
larger data sets for effective higher-order network reconstruction,
especially in the presence of noise. Notably, GLOC requires
the fewest additional observations while maintaining
a strong overall performance, confirming that the method
is appropriate for higher-order interactions even in
noisy environments, and highlighting its robustness.

\subsection{Empirical complexes}
To validate the general applicability of our methods,
we further test them on empirical simplicial
complexes. Specifically, we build the complexes from
the Malawi, SFHH and LH10 data sets, as described in
Subsection~\ref{Sec:Empirical}, and reconstruct the networks
from observations of the Prisoner's Dilemma game.

The values of the metrics for our reconstructions
are reported in Table~S4 and S5 of the Supplementary Material. When noise is
introduced in the payoff matrix, all of three methods perform
well in reconstructing 1-simplices. However, GLO and GLOC
consistently outperform PBP in 2-simplex reconstruction,
demonstrating their superior ability to capture higher-order interactions. After the addition of noise in payoffs,
the difference in performance is amplified, and GLOC
remains the only method that maintains high scores
for all metrics. Performance differences become more
pronounced when noise contaminates the information of
strategies. As seen in Table S5, despite the notable
performance declines that all of three methods suffer
under small perturbations in strategy observations ($\psi$ = 1\%),
GLOC still reaches the highest $F_1$ scores to reconstruct
2-simplices. These results highlight GLOC's robustness and
superiority, ensuring high accuracy even when empirical data
is subject to fluctuations and uncertainty during transient phases.

Measurements of the $F_1$~score for different
sizes of the observation set are shown in Fig.~\ref{Fig:Emp_F1}
and in Fig.~\ref{Fig:Emp_F1_SIG1} in the absence
and in the presence of noise regarding payoffs, respectively.
In all cases, and consistently with the results
obtained on synthetic complexes, GLOC is the
method that requires the smallest number of observations~$M$
to reach a high quality of reconstruction, on
both 1-simplices and 2-simplices. In fact, the
number of required observations is only around~15,
which is an even smaller value than the one needed
for synthetic simplicial complexes. As illustrated in Fig.~S2 the impact on reconstruction further
varies across methods when strategy noise ($\psi$ = 1\%)
is introduced. PBP, while relatively stable, requires a
large number of observations to reach a
reasonable $F_1$ score. GLO, on the other hand, shows
extreme variability, with large fluctuations and poor
performance across different network types, indicating
its sensitivity to erroneous strategies. In contrast, GLOC
remains the most reliable method, where highest $F_1$ scores
can be guaranteed within the smallest observation data sets.
The robustness of GLOC against noise highlights its effectiveness
in reconstructing simplicial complexes using data captured
exclusively during the transient phase of evolutionary game
dynamics, rather than relying on equilibrium states. This
capability is particularly valuable in real-world scenarios
where networked interactions evolve dynamically, and equilibrium
conditions are rarely observed. Moreover, GLOC’s ability to
maintain relatively high accuracy with limited, noisy data
suggests that it can significantly reduce the amount
of data collected in scenarios where resources
are limited and where data contamination is unavoidable
during collection.

\section{Conclusions}\label{Sec:Discussion}
In summary, we have introduced three methods
for the reconstruction of simplicial complexes.
One is a local point-by-point estimator~(PBP),
and the other two are a global approach~(GLO)
and a global estimator which explicitly includes
simplicial constraints~(GLOC). These are amongst
the very few approaches that can reconstruct
higher-order networks using observations of
evolutionary dynamics in transient phases.
All three methods perform
well according to a number of different metrics
when accounting for the reconstructed simple
edges. However, while still obtaining high-quality
results, PBP does not provide a perfect accuracy
when measuring the reconstructed 2-simplices.
In this latter case, GLO and GLOC furnish more precise
results, with GLOC being particularly useful
in the case of noisy data. Additionally, GLOC achieves highly precise
results with significantly fewer observations,
regardless of whether the payoff information is
contaminated or the strategies deviate from
equilibrium due to noise.

These characteristics of the behaviour of the
methods are independent of the origin of the
network being reconstructed. In fact, the superiority of~GLOC remains unchanged whether
different methods are applied to synthetic models of simplicial
complexes, or to higher-order structures derived
from empirical data.

This suggests several considerations about those scenarios where each
of the three methods is the most appropriate for being used. For instance, in situations where data
collection is not an expensive task, and where
a very high quality of reconstruction of higher-order
interactions, yet short of perfection, is acceptable,
PBP would be the method of choice, as it is
computationally less complex, with the cost
of requiring more observation data. If, instead
the reconstruction quality is of absolute importance,
or if the costs associated to the process of
data gathering are higher than those related
to computing time, GLO is the best choice.
However, if one cannot be sure that the uncertainties
on the data are negligible, while still requiring
the best possible reconstruction regardless
of the complexity of the task, then GLOC is
the method to use. Our methods are
rooted in mathematical optimization, providing a
flexible and adaptive framework for reconstructing
higher-order structures in real-world distributed
systems, where equilibrium is often impractical due
to the dynamic and fluctuating nature of interactions.
By leveraging transient-phase data rather than relying
on equilibrium states, our approach offers a more
realistic and effective solution for studying higher-order
interactions within simplicial complexes. This framework
fills a critical methodological gap by systematically
addressing the challenges of reconstructing higher-order evolutionary
game dynamics, a perspective that has not been thoroughly
explored in previous research. Furthermore, our suite of
optimization-based methods allows for tailored reconstruction
strategies depending on specific system requirements, while
avoiding the intrinsic limitations of traditional
statistical techniques that often assume stationarity
or long-term convergence.


\acknowledgments
Y.M. and Z.J. acknowledge partial support from the National Natural
Science Foundation of China (No.~72171084 and No.~91746108) and the
China Scholar Council State Scholarship Fund (No. 202306740033). F.F.
acknowledges partial support from the China Scholar Council State Scholarship
Fund (No.~202306830151). C.I.D.G. acknowledges funding from the Bulgarian
Ministry of Education and Science, under project number BG-RRP-2.004-0006-C02.
S.B. acknowledges support from the project n.PGR01177 ``Systems with
higher order interactions and fractional derivatives for applications
to AI and high performance computing'' of the Italian Ministry of Foreign
Affairs and International Cooperation.

\clearpage

\end{document}


\title{Reconstructing simplicial complexes from evolutionary games \\ Supplementary Material}
\author{Yin-Jie Ma}
    \affiliation{School of Business, East China University of Science and Technology, Shanghai, China}
    \affiliation{Research Center for Econophysics, East China University of Science and Technology, Shanghai, China}
    \affiliation{CNR - Institute of Complex Systems, Via Madonna del Piano 10, I-50019, Sesto Fiorentino, Italy}
\author{Zhi-Qiang Jiang}\email{Corresponding author: zqjiang@ecust.edu.cn}
    \affiliation{School of Business, East China University of Science and Technology, Shanghai, China}
    \affiliation{Research Center for Econophysics, East China University of Science and Technology, Shanghai, China}
\author{Fanshu Fang}
    \affiliation{CNR - Institute of Complex Systems, Via Madonna del Piano 10, I-50019, Sesto Fiorentino, Italy}
    \affiliation{College of Economics and Management, Nanjing University of Aeronautics and Astronautics, Nanjing, China}
\author{Charo I. \surname{del Genio}}
    \affiliation{Institute of Smart Agriculture for Safe and Functional Foods and Supplements, Trakia University, Stara Zagora 6000, Bulgaria}
    \affiliation{Research Institute of Interdisciplinary Intelligent Science, Ningbo University of Technology, 315104, Ningbo, China}
\author{Stefano Boccaletti}\email{Corresponding author: stefano.boccaletti@gmail.com}
    \affiliation{CNR - Institute of Complex Systems, Via Madonna del Piano 10, I-50019, Sesto Fiorentino, Italy}
    \affiliation{Research Institute of Interdisciplinary Intelligent Science, Ningbo University of Technology, 315104, Ningbo, China}
    \affiliation{Sino-Europe Complexity Science Center, School of Mathematics, North University of China, 030051, Taiyuan, China}

\date{\today}

\maketitle

\begin{longtable}{@{}c@{\qquad}c@{\qquad}c@{}}
\toprule
Method & \multicolumn{2}{c}{ACC/TPR/PPV/TNR/$F_1$/MCC/AUROC/AUPR} \\
\midrule

(Net type, Game type, $d$) & (NPA, PD, 1) & (NPA, PD, 2)\\
PBP & 0.99/0.96/1.00/1.00/0.98/0.97/0.98/0.98 & 1.00/0.92/0.51/1.00/0.64/0.67/0.96/0.71\\
GLO & 0.99/1.00/0.95/0.99/0.97/0.97/1.00/0.98 & 1.00/1.00/0.98/1.00/0.99/0.99/1.00/0.99\\
GLOC & 1.00/1.00/0.99/1.00/1.00/1.00/1.00/1.00 & 1.00/1.00/0.99/1.00/1.00/1.00/1.00/1.00\\

(Net type, Game type, $d$) & (PA, PD, 1) & (PA, PD, 2)\\
PBP & 0.96/0.76/1.00/1.00/0.86/0.85/0.88/0.90 & 0.99/0.58/0.37/1.00/0.45/0.46/0.79/0.48\\
GLO & 1.00/1.00/0.97/0.99/0.98/0.98/1.00/0.98 & 1.00/1.00/0.98/1.00/0.99/0.99/1.00/0.99\\
GLOC & 1.00/1.00/1.00/1.00/1.00/1.00/1.00/1.00 & 1.00/1.00/0.99/1.00/1.00/1.00/1.00/1.00\\

(Net type, Game type, $d$) & (MIX, PD, 1) & (MIX, PD, 2)\\
PBP & 0.89/0.26/1.00/1.00/0.41/0.48/0.63/0.69 & 0.99/0.04/0.39/1.00/0.07/0.12/0.52/0.22\\
GLO & 1.00/1.00/0.99/1.00/0.99/0.99/1.00/0.99 & 1.00/1.00/0.96/1.00/0.98/0.98/1.00/0.98\\
GLOC & 1.00/1.00/1.00/1.00/1.00/1.00/1.00/1.00 & 1.00/1.00/0.98/1.00/0.99/0.99/1.00/0.99\\
\midrule

(Net type, Game type, $d$) & (NPA, SD, 1) & (NPA, SD, 2)\\
PBP & 0.99/0.93/1.00/1.00/0.96/0.96/0.97/0.97 & 0.99/0.83/0.38/0.99/0.52/0.56/0.91/0.61\\
GLO & 1.00/1.00/0.99/1.00/0.99/0.99/1.00/0.99 & 1.00/1.00/1.00/1.00/1.00/1.00/1.00/1.00\\
GLOC & 1.00/1.00/0.98/1.00/0.99/0.99/1.00/0.99 & 1.00/1.00/0.97/1.00/0.98/0.98/1.00/0.98\\

(Net type, Game type, $d$) & (PA, SD, 1) & (PA, SD, 2)\\
PBP & 0.96/0.74/1.00/1.00/0.85/0.84/0.87/0.89 & 0.99/0.59/0.32/0.99/0.41/0.43/0.79/0.45\\
GLO & 1.00/1.00/0.99/1.00/1.00/0.99/1.00/1.00 & 1.00/1.00/0.98/1.00/0.99/0.99/1.00/0.99\\
GLOC & 1.00/1.00/0.98/1.00/0.99/0.99/1.00/0.99 & 1.00/1.00/0.99/1.00/0.99/0.99/1.00/0.99\\

(Net type, Game type, $d$) & (MIX, SD, 1) & (MIX, SD, 2)\\
PBP & 0.89/0.30/0.96/1.00/0.46/0.50/0.65/0.69 & 0.99/0.13/0.38/1.00/0.19/0.21/0.57/0.26\\
GLO & 1.00/1.00/0.99/1.00/1.00/1.00/1.00/1.00 & 1.00/1.00/0.96/1.00/0.98/0.98/1.00/0.98\\
GLOC & 1.00/1.00/0.99/1.00/1.00/1.00/1.00/1.00 & 1.00/1.00/0.99/1.00/1.00/1.00/1.00/1.00\\
\midrule

(Net type, Game type, $d$) & (NPA, SH, 1) & (NPA, SH, 2)\\
PBP & 0.99/0.97/1.00/1.00/0.98/0.98/0.98/0.98 & 0.99/0.96/0.47/1.00/0.63/0.67/0.98/0.71\\
GLO & 1.00/1.00/0.99/1.00/1.00/0.99/1.00/1.00 & 1.00/1.00/0.93/1.00/0.96/0.96/1.00/0.96\\
GLOC & 1.00/1.00/0.97/0.99/0.99/0.98/1.00/0.99 & 1.00/1.00/0.99/1.00/0.99/0.99/1.00/0.99\\

(Net type, Game type, $d$) & (PA, SH, 1) & (PA, SH, 2)\\
PBP & 0.99/0.94/1.00/1.00/0.97/0.96/0.97/0.97 & 0.99/0.94/0.40/0.99/0.56/0.61/0.97/0.67\\
GLO & 1.00/1.00/1.00/1.00/1.00/1.00/1.00/1.00 & 1.00/1.00/1.00/1.00/1.00/1.00/1.00/1.00\\
GLOC & 1.00/1.00/1.00/1.00/1.00/1.00/1.00/1.00 & 1.00/1.00/0.99/1.00/0.99/0.99/1.00/0.99\\

(Net type, Game type, $d$) & (MIX, SH, 1) & (MIX, SH, 2)\\
PBP & 0.90/0.36/0.99/1.00/0.52/0.56/0.68/0.72 & 0.99/0.09/0.31/1.00/0.14/0.16/0.55/0.20\\
GLO & 1.00/1.00/0.99/1.00/1.00/1.00/1.00/1.00 & 1.00/1.00/0.99/1.00/0.99/0.99/1.00/0.99\\
GLOC & 1.00/1.00/0.99/1.00/0.99/0.99/1.00/0.99 & 1.00/1.00/1.00/1.00/1.00/1.00/1.00/1.00\\
\midrule
\newpage
(Net type, Game type, $d$) & (NPA, H, 1) & (NPA, H, 2)\\
PBP & 0.99/0.91/1.00/1.00/0.95/0.95/0.96/0.96 & 1.00/0.82/0.62/1.00/0.70/0.71/0.91/0.72\\
GLO & 1.00/1.00/1.00/1.00/1.00/1.00/1.00/1.00 & 1.00/0.99/1.00/1.00/0.99/0.99/0.99/0.99\\
GLOC & 1.00/1.00/1.00/1.00/1.00/1.00/1.00/1.00 & 1.00/1.00/0.99/1.00/0.99/0.99/1.00/0.99\\

(Net type, Game type, $d$) & (PA, H, 1) & (PA, H, 2)\\
PBP & 0.97/0.80/0.99/1.00/0.89/0.88/0.90/0.91 & 1.00/0.70/0.57/1.00/0.61/0.62/0.85/0.63\\
GLO & 1.00/1.00/1.00/1.00/1.00/1.00/1.00/1.00 & 1.00/0.97/0.94/1.00/0.96/0.96/0.99/0.96\\
GLOC & 1.00/1.00/1.00/1.00/1.00/1.00/1.00/1.00 & 1.00/1.00/0.97/1.00/0.98/0.98/1.00/0.98\\

(Net type, Game type, $d$) & (MIX, H, 1) & (MIX, H, 2)\\
PBP & 0.89/0.29/1.00/1.00/0.44/0.50/0.64/0.70 & 0.99/0.10/0.33/1.00/0.15/0.17/0.55/0.21\\
GLO & 1.00/1.00/0.99/0.99/0.99/0.99/1.00/1.00 & 1.00/0.65/0.98/1.00/0.65/0.67/0.83/0.82\\
GLOC & 1.00/1.00/1.00/1.00/1.00/1.00/1.00/1.00 & 1.00/1.00/0.85/1.00/0.91/0.91/1.00/0.92\\
\bottomrule\\
\caption{\textbf{The GLO and GLOC methods perform better than PBP.}
Values of accuracy~(ACC), true positive rate~(TPR), positive predictive
value~(PPV), true negative rate~(TNR), $F_1$ score, Matthews correlation
coefficient~(MCC), as well as the area under the receiver operating
characteristic curve~(AUROC) and that under the precision-recall curve~(AUPR)
show that all three methods perform well in the reconstruction of 1-simplices.
However, GLO and GLOC provide much better results on 2-simplices, especially
according to more complex metrics. The results are reported by type
of synthetic simplicial complex model (NPA, PA and MIX), type of evolutionary
game (Prisoner's Dilemma -- PD, Snowdrift -- SD, Stag Hunt -- SH and Harmony -- H),
and dimension~$d$ of the simplices considered.
In all cases, the number of nodes is~50, the number of observations is~95
and $\rho=0.5$.}
\end{longtable}

\begin{table}[h]
\centering
\begin{tabular}{@{}c@{\qquad}c@{\qquad}c@{}}
\toprule
Method & \multicolumn{2}{c}{ACC/TPR/PPV/TNR/$F_1$/MCC/AUROC/AUPR} \\
\midrule

(Net type, $q$, $d$) & (NPA, $1\%$, 1) & (NPA, $1\%$, 2)\\
PBP & 0.99/0.90/1.00/1.00/0.95/0.94/0.95/0.96 & 0.99/0.79/0.42/1.00/0.54/0.57/0.89/0.61\\
GLO & 1.00/1.00/0.99/1.00/1.00/1.00/1.00/1.00 & 1.00/1.00/0.99/1.00/1.00/1.00/1.00/1.00\\
GLOC & 1.00/1.00/0.99/1.00/1.00/1.00/1.00/1.00 & 1.00/1.00/1.00/1.00/1.00/1.00/1.00/1.00\\

(Net type, $q$, $d$) & (NPA, $5\%$, 1) & (NPA, $5\%$, 2)\\
PBP & 0.98/0.89/1.00/1.00/0.94/0.93/0.95/0.95 & 0.99/0.78/0.37/0.99/0.50/0.53/0.89/0.58\\
GLO & 1.00/1.00/0.98/1.00/0.99/0.99/1.00/0.99 & 1.00/0.96/0.99/1.00/0.98/0.98/0.98/0.98\\
GLOC & 1.00/1.00/0.99/1.00/0.99/0.99/1.00/0.99 & 1.00/1.00/0.98/1.00/0.99/0.99/1.00/0.99\\
\midrule

(Net type, $q$, $d$) & (PA, $1\%$, 1) & (PA, $1\%$, 2)\\
PBP & 0.97/0.80/1.00/1.00/0.89/0.88/0.90/0.92 & 0.99/0.66/0.34/0.99/0.45/0.47/0.83/0.50\\
GLO & 1.00/1.00/1.00/1.00/1.00/1.00/1.00/1.00 & 1.00/1.00/0.98/1.00/0.99/0.99/1.00/0.99\\
GLOC & 1.00/1.00/0.99/1.00/1.00/0.99/1.00/1.00 & 1.00/1.00/0.98/1.00/0.99/0.99/1.00/0.99\\

(Net type, $q$, $d$) & (PA, $5\%$, 1) & (PA, $5\%$, 2)\\
PBP & 0.96/0.77/1.00/1.00/0.87/0.86/0.88/0.90 & 0.99/0.64/0.35/0.99/0.45/0.47/0.82/0.49\\
GLO & 0.91/0.41/0.98/1.00/0.41/0.45/0.70/0.74 & 1.00/0.26/0.60/1.00/0.30/0.20/0.63/0.63\\
GLOC & 0.97/0.80/0.98/1.00/0.91/0.81/0.90/0.91 & 1.00/0.80/0.95/1.00/0.78/0.79/0.90/0.88\\
\midrule

(Net type, $q$, $d$) & (MIX, $1\%$, 1) & (MIX, $1\%$, 2)\\
PBP & 0.89/0.28/0.99/1.00/0.44/0.50/0.64/0.69 & 0.99/0.06/0.26/1.00/0.09/0.12/0.53/0.16\\
GLO & 1.00/1.00/1.00/1.00/1.00/1.00/1.00/1.00 & 1.00/1.00/1.00/1.00/1.00/1.00/1.00/1.00\\
GLOC & 1.00/1.00/0.99/1.00/1.00/1.00/1.00/1.00 & 1.00/1.00/1.00/1.00/1.00/0.99/1.00/0.99\\

(Net type, $q$, $d$) & (MIX, $5\%$, 1) & (MIX, $5\%$, 2)\\
PBP & 0.88/0.23/0.97/1.00/0.37/0.43/0.62/0.66 & 0.99/0.06/0.15/1.00/0.08/0.09/0.53/0.11\\
GLO & 0.85/0.02/1.00/1.00/0.05/0.14/0.51/0.59 & 0.99/0.00/0.00/1.00/0.00/UD/0.50/0.50\\
GLOC & 0.91/0.38/1.00/1.00/0.41/0.46/0.69/0.74 & 1.00/0.35/0.58/1.00/0.37/UD/0.68/0.67\\
\bottomrule
\end{tabular}
\caption{\textbf{The GLOC method performs consistently better than PBP and GLO when 
the measured payoff matrix is contaminated by noise}.
Values of accuracy~(ACC), true positive rate~(TPR), positive predictive
value~(PPV), true negative rate~(TNR), $F_1$ score, Matthews correlation
coefficient~(MCC), as well as the area under the receiver operating
characteristic curve~(AUROC) and that under the precision-recall curve~(AUPR)
for noisy data from a Prisoner's Dilemma game show that, while PBP and GLO
are affected by higher levels of noise in payoffs, GLOC maintains high scores across
the different metrics. The results are reported by type of synthetic simplicial
complex model (NPA, PA and MIX), noise levels in payoffs ($q=1\%$ or $5\%$), and
dimension~$d$ of the simplices considered. In all cases, the number of
nodes is~50, the number of observations is~95 and $\rho=0.5$. Note that
in two cases, there were no true positive nor false positives; this led
to the denominator of the expression for the MCC to vanish, and the MCC
itself being undefined. We have marked these two cases as UD.
}
\end{table}

\begin{table}[h]
\centering
\begin{tabular}{@{}c@{\qquad}c@{\qquad}c@{}}
\toprule
Method & \multicolumn{2}{c}{ACC/TPR/PPV/TNR/$F_1$/MCC/AUROC/AUPR} \\
\midrule

(Net type, $\psi$, $d$) & (NPA, $1\%$, 1) & (NPA, $1\%$, 2)\\
PBP & 0.92/0.52/0.91/0.99/0.66/0.65/0.75/0.75 & 0.98/0.20/0.07/0.99/0.10/0.11/0.60/0.14\\
GLO & 0.85/0.04/0.41/0.99/0.07/0.09/0.52/0.30 & 0.99/0.04/0.02/0.99/0.03/0.02/0.52/0.03\\
GLOC & 0.99/0.99/0.95/0.99/0.97/0.96/0.99/0.97 & 0.99/0.99/0.30/0.99/0.46/0.54/0.99/0.64\\
\midrule

(Net type, $\psi$, $d$) & (NPA, $5\%$, 1) & (NPA, $5\%$, 2)\\
PBP & 0.91/0.48/0.87/0.99/0.56/0.58/0.73/0.71 & 0.98/0.34/0.08/0.99/0.13/0.16/0.67/0.21\\
GLO & 0.85/0.06/0.52/0.99/0.10/0.14/0.52/0.36 & 0.99/0.07/0.03/0.99/0.04/0.04/0.53/0.05\\
GLOC & 0.99/0.99/0.95/0.99/0.97/0.96/0.99/0.97 & 0.99/0.99/0.30/0.99/0.46/0.54/0.99/0.64\\
\midrule

(Net type, $\psi$, $d$) & (PA, $1\%$, 1) & (PA, $1\%$, 2)\\
PBP & 0.91/0.43/0.90/0.99/0.58/0.58/0.71/0.71 & 0.98/0.14/0.05/0.99/0.07/0.08/0.56/0.10\\
GLO & 0.85/0.04/0.45/0.99/0.07/0.09/0.51/0.31 & 0.99/0.01/0.01/0.99/0.01/0.00/0.50/0.01\\
GLOC & 0.99/0.99/0.95/0.99/0.97/0.96/0.99/0.97 & 0.99/0.99/0.31/0.99/0.47/0.55/0.99/0.65\\
\midrule

(Net type, $\psi$, $d$) & (PA, $5\%$, 1) & (PA, $5\%$, 2)\\
PBP & 0.87/0.23/0.80/0.99/0.35/0.38/0.61/0.57 & 0.99/0.04/0.02/0.99/0.02/0.02/0.51/0.03\\
GLO & 0.63/0.40/0.46/0.67/0.20/0.15/0.53/0.47 & 0.99/0.04/0.02/0.99/0.03/0.02/0.52/0.03\\
GLOC & 0.94/0.66/0.65/0.99/0.64/0.62/0.82/0.68 & 0.99/0.66/0.20/0.99/0.31/0.36/0.83/0.43\\
\midrule

(Net type, $\psi$, $d$) & (MIX, $1\%$, 1) & (MIX, $1\%$, 2)\\
PBP & 0.87/0.21/0.82/0.99/0.33/0.37/0.60/0.57 & 0.98/0.04/0.03/0.99/0.03/0.03/0.52/0.04\\
GLO & 0.89/0.34/0.50/0.99/0.35/0.35/0.67/0.47 & 0.99/0.34/0.14/0.99/0.18/0.21/0.67/0.24\\
GLOC & 0.99/0.99/0.96/0.99/0.97/0.97/0.99/0.97 & 0.99/0.99/0.41/0.99/0.47/0.63/0.99/0.70\\
\midrule

(Net type, $\psi$, $d$) & (MIX, $5\%$, 1) & (MIX, $5\%$, 2)\\
PBP & 0.87/0.16/0.78/0.99/0.27/0.31/0.58/0.53 & 0.98/0.02/0.02/0.99/0.02/0.01/0.51/0.02\\
GLO & 0.90/0.36/0.65/0.99/0.38/0.39/0.67/0.55 & 0.99/0.34/0.14/0.99/0.18/0.21/0.66/0.24\\
GLOC & 0.89/0.35/0.55/0.99/0.36/0.36/0.67/0.50 & 0.99/0.33/0.14/0.99/0.20/0.21/0.66/0.24\\
\bottomrule
\end{tabular}
\caption{\textbf{The GLOC method performs consistently better than PBP and GLO when the measured strategy matrix is contaminated by noise.}
Values of accuracy~(ACC), true positive rate~(TPR), positive predictive
value~(PPV), true negative rate~(TNR), $F_1$ score, Matthews correlation
coefficient~(MCC), as well as the area under the receiver operating
characteristic curve~(AUROC) and that under the precision-recall curve~(AUPR) 
for noisy data from a Prisoner's Dilemma game show that, despite sharp drops across different metrics compared with those in contaminated payoffs, 
GLOC still reaches the highest ones especially when tiny noise exists in 
strategies. The results are reported by the type of synthetic simplicial 
complex model, amount of noise $\psi$ in strategies and dimension $d$ of the simplices considered.
}
\end{table}

\begin{table}[h]
\centering
\begin{tabular}{@{}c@{\qquad}c@{\qquad}c@{}}
\toprule
Method & \multicolumn{2}{c}{ACC/TPR/PPV/TNR/$F_1$/MCC/AUROC/AUPR} \\
\midrule

(Net type, $q$, $d$) & (Malawi, $0\%$, 1) & (Malawi, $0\%$, 2)\\
PBP & 1.00/1.00/0.99/1.00/1.00/1.00/1.00/1.00 & 1.00/1.00/0.69/1.00/0.81/0.83/1.00/0.84\\
GLO & 1.00/1.00/0.99/1.00/0.99/0.99/1.00/0.99 & 1.00/1.00/0.96/1.00/0.98/0.98/1.00/0.98\\
GLOC & 0.99/1.00/0.91/0.99/0.95/0.95/1.00/0.95 & 1.00/1.00/0.92/1.00/0.96/0.96/1.00/0.96\\

(Net type, $q$, $d$) & (Malawi, $1\%$, 1) & (Malawi, $1\%$, 2)\\
PBP & 1.00/1.00/1.00/1.00/1.00/1.00/1.00/1.00 & 1.00/1.00/0.77/1.00/0.87/0.88/1.00/0.88\\
GLO & 1.00/1.00/0.97/1.00/0.99/0.99/1.00/0.99 & 1.00/1.00/0.98/1.00/0.99/0.99/1.00/0.99\\
GLOC & 1.00/1.00/0.99/1.00/0.99/0.99/1.00/0.99 & 1.00/1.00/0.97/1.00/0.99/0.99/1.00/0.99\\

(Net type, $q$, $d$) & (Malawi, $5\%$, 1) & (Malawi, $5\%$, 2)\\
PBP & 1.00/1.00/1.00/1.00/1.00/1.00/1.00/1.00 & 1.00/1.00/0.73/1.00/0.84/0.86/1.00/0.87\\
GLO & 1.00/1.00/0.95/1.00/0.97/0.97/1.00/0.98 & 1.00/1.00/1.00/1.00/1.00/1.00/1.00/1.00\\
GLOC & 1.00/1.00/0.98/1.00/0.99/0.99/1.00/0.99 & 1.00/1.00/0.99/1.00/1.00/1.00/1.00/1.00\\
\midrule

(Net type, $q$, $d$) & (SFHH, $0\%$, 1) & (SFHH, $0\%$, 2)\\
PBP & 1.00/1.00/1.00/1.00/1.00/1.00/1.00/1.00 & 1.00/1.00/0.71/1.00/0.83/0.84/1.00/0.85\\
GLO & 1.00/1.00/0.96/1.00/0.98/0.98/1.00/0.98 & 1.00/1.00/1.00/1.00/1.00/1.00/1.00/1.00\\
GLOC & 1.00/1.00/0.99/1.00/1.00/1.00/1.00/1.00 & 1.00/1.00/0.99/1.00/0.99/0.99/1.00/0.99\\

(Net type, $q$, $d$) & (SFHH, $1\%$, 1) & (SFHH, $1\%$, 2)\\
PBP & 1.00/1.00/0.99/1.00/1.00/1.00/1.00/1.00 & 1.00/1.00/0.81/1.00/0.89/0.90/1.00/0.90\\
GLO & 1.00/1.00/0.98/1.00/0.99/0.99/1.00/0.99 & 1.00/1.00/0.98/1.00/0.99/0.99/1.00/0.99\\
GLOC & 1.00/1.00/0.99/1.00/0.99/0.99/1.00/0.99 & 1.00/1.00/1.00/1.00/1.00/1.00/1.00/1.00\\

(Net type, $q$, $d$) & (SFHH, $5\%$, 1) & (SFHH, $5\%$, 2)\\
PBP & 1.00/0.98/1.00/1.00/0.99/0.99/0.99/0.99 & 1.00/0.99/0.73/1.00/0.84/0.85/0.99/0.86\\
GLO & 1.00/1.00/0.94/1.00/0.97/0.97/1.00/0.97 & 1.00/1.00/1.00/1.00/1.00/1.00/1.00/1.00\\
GLOC & 1.00/1.00/0.99/1.00/1.00/1.00/1.00/1.00 & 1.00/1.00/0.97/1.00/0.99/0.99/1.00/0.99\\
\midrule

(Net type, $q$, $d$) & (LH10, $0\%$, 1) & (LH10, $0\%$, 2)\\
PBP & 1.00/0.99/1.00/1.00/0.99/0.99/1.00/0.99 & 1.00/1.00/0.64/1.00/0.78/0.80/1.00/0.82\\
GLO & 0.99/1.00/0.94/0.99/0.97/0.96/1.00/0.97 & 1.00/1.00/0.99/1.00/0.99/0.99/1.00/0.99\\
GLOC & 1.00/1.00/0.99/1.00/0.99/0.99/1.00/0.99 & 1.00/1.00/0.98/1.00/0.99/0.99/1.00/0.99\\

(Net type, $q$, $d$) & (LH10, $1\%$, 1) & (LH10, $1\%$, 2)\\
PBP & 1.00/0.98/1.00/1.00/0.99/0.99/0.99/0.99 & 1.00/0.98/0.77/1.00/0.86/0.87/0.99/0.87\\
GLO & 0.99/1.00/0.95/0.99/0.97/0.97/1.00/0.97 & 1.00/1.00/1.00/1.00/1.00/1.00/1.00/1.00\\
GLOC & 1.00/1.00/0.99/1.00/1.00/0.99/1.00/1.00 & 1.00/1.00/0.99/1.00/1.00/1.00/1.00/1.00\\

(Net type, $q$, $d$) & (LH10, $5\%$, 1) & (LH10, $5\%$, 2)\\
PBP & 0.97/0.77/1.00/1.00/0.87/0.86/0.88/0.90 & 1.00/0.61/0.70/1.00/0.65/0.65/0.80/0.65\\
GLO & 1.00/1.00/0.99/1.00/0.99/0.99/1.00/0.99 & 1.00/0.73/1.00/1.00/0.75/0.77/0.87/0.87\\
GLOC & 1.00/1.00/0.99/1.00/1.00/1.00/1.00/1.00 & 1.00/1.00/0.99/1.00/1.00/1.00/1.00/1.00\\
\bottomrule
\end{tabular}
\caption{\textbf{The GLO and GLOC methods perform better than PBP on empirical complexes when
the measured payoff matrix is contaminated by noise}.Values of accuracy~(ACC), true positive rate~(TPR), 
positive predictive value~(PPV), true
negative rate~(TNR), $F_1$ score, Matthews correlation coefficient~(MCC), as well as the area
under the receiver operating characteristic curve~(AUROC) and that under the precision-recall
curve~(AUPR) show that all three methods perform well in the reconstruction of 1-simplices
without noise. However, GLO and GLOC outperform PBP on 2-simplices. This difference is amplified
when noise is added to the data, in which case GLOC is the only method that maintains a good
performance. The results are reported by data set, amount of noise~$q$ and dimension~$d$
of the simplices considered.
}
\end{table}

\begin{table}[h]
\centering
\begin{tabular}{@{}c@{\qquad}c@{\qquad}c@{}}
\toprule
Method & \multicolumn{2}{c}{ACC/TPR/PPV/TNR/$F_1$/MCC/AUROC/AUPR} \\
\midrule

(Net type, $\psi$, $d$) & (Malawi, $1\%$, 1) & (Malawi, $1\%$, 2)\\
PBP & 0.99/0.91/0.87/0.99/0.89/0.88/0.95/0.90 & 0.99/0.79/0.11/0.99/0.20/0.29/0.89/0.45\\
GLO & 0.99/0.93/0.87/0.99/0.90/0.89/0.96/0.90 & 0.99/0.22/0.04/0.99/0.06/0.08/0.60/0.13\\
GLOC & 0.99/0.96/0.88/0.99/0.92/0.91/0.97/0.92 & 0.99/0.99/0.15/0.99/0.25/0.38/0.99/0.57\\
\midrule

(Net type, $\psi$, $d$) & (Malawi, $5\%$, 1) & (Malawi, $5\%$, 2)\\
PBP & 0.96/0.56/0.80/0.99/0.66/0.65/0.77/0.70 & 0.99/0.25/0.04/0.99/0.07/0.10/0.62/0.15\\
GLO & 0.93/0.14/0.49/0.99/0.22/0.24/0.57/0.35 & 0.99/0.01/0.00/0.99/0.00/0.00/0.50/0.01\\
GLOC & 0.77/0.35/0.48/0.80/0.26/0.25/0.58/0.44 & 0.99/0.41/0.06/0.99/0.11/0.16/0.70/0.24\\
\midrule

(Net type, $\psi$, $d$) & (SFHH, $1\%$, 1) & (SFHH, $1\%$, 2)\\
PBP & 0.98/0.77/0.85/0.99/0.80/0.79/0.88/0.82 & 0.99/0.51/0.11/0.99/0.18/0.23/0.75/0.31\\
GLO & 0.94/0.24/0.44/0.99/0.26/0.28/0.61/0.36 & 0.99/0.06/0.01/0.99/0.02/0.02/0.52/0.04\\
GLOC & 0.98/0.82/0.85/0.99/0.83/0.82/0.90/0.84 & 0.99/0.96/0.20/0.99/0.33/0.43/0.97/0.58\\
\midrule

(Net type, $\psi$, $d$) & (SFHH, $5\%$, 1) & (SFHH, $5\%$, 2)\\
PBP & 0.95/0.41/0.75/0.99/0.53/0.53/0.70/0.60 & 0.99/0.07/0.02/0.99/0.03/0.03/0.53/0.04\\
GLO & 0.92/0.16/0.25/0.98/0.15/0.14/0.57/0.23 & 0.99/0.01/0.00/0.99/0.00/0.00/0.50/0.01\\
GLOC & 0.94/0.26/0.65/0.99/0.36/0.38/0.62/0.48 & 0.99/0.38/0.09/0.99/0.14/0.18/0.68/0.23\\
\midrule

(Net type, $\psi$, $d$) & (LH10, $1\%$, 1) & (LH10, $1\%$, 2)\\
PBP & 0.93/0.47/0.86/0.99/0.61/0.61/0.73/0.69 & 0.98/0.20/0.10/0.99/0.13/0.13/0.59/0.15\\
GLO & 0.89/0.04/0.34/0.99/0.07/0.08/0.52/0.24 & 0.98/0.01/0.01/0.99/0.01/0.00/0.50/0.01\\
GLOC & 0.95/0.63/0.88/0.99/0.72/0.72/0.81/0.78 & 0.99/0.77/0.32/0.99/0.45/0.49/0.88/0.54\\
\midrule

(Net type, $\psi$, $d$) & (LH10, $5\%$, 1) & (LH10, $5\%$, 2)\\
PBP & 0.92/0.39/0.78/0.99/0.48/0.50/0.69/0.62 & 0.98/0.21/0.07/0.99/0.11/0.12/0.60/0.15\\
GLO & 0.88/0.03/0.25/0.99/0.05/0.05/0.51/0.19 & 0.98/0.01/0.01/0.99/0.01/0.00/0.50/0.01\\
GLOC & 0.94/0.99/0.11/0.03/0.20/0.05/0.51/0.55 & 0.98/0.19/0.10/0.99/0.13/0.13/0.59/0.14\\
\bottomrule
\end{tabular}
\caption{\textbf{GLOC performs the best to reconstruct 
empirical complexes when tiny noise exists in 
a strategy matrix.} 
Values of accuracy~(ACC), true positive rate~(TPR), positive predictive
value~(PPV), true negative rate~(TNR), $F_1$ score, Matthews correlation
coefficient~(MCC), as well as the area under the receiver operating
characteristic curve~(AUROC) and that under the precision-recall curve~(AUPR)
show that GLOC consistently outperforms the other two methods to
reconstruct 2-simplices when a
minority part ($\psi$ = 1 \%) of strategies have been mistaken into the opposite ones. 
The results are reported by data set, 
amount of noise $\psi$ in strategies and dimension $d$ of the simplices considered.
}
\end{table}

\begin{figure*}[t]
\centering
\includegraphics[width=0.95\textwidth]{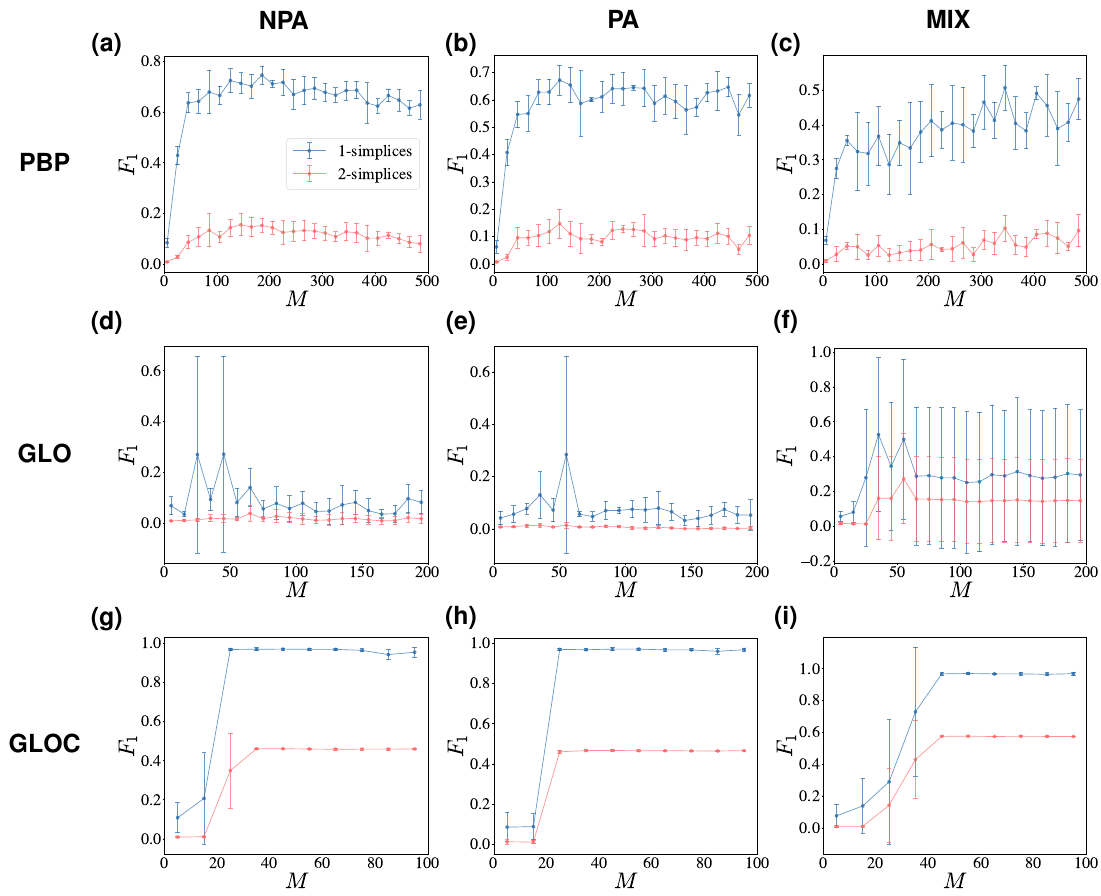}
\caption{\textbf{GLOC outperforms PBP and GLO for 
noisy data in strategies.} In sharp contrast to the results of 
contaminated payoffs (Fig.~3), the performance 
of all three methods on noisy data ($\psi$ = 1\%) of strategies from a
Prisoner’s Dilemma game drops. However, GLOC still reaches the highest $F_1$ scores
to reconstruct both 1-simplices (blue) and 2-simplices (red) in all of three synthetic 
simplicial complex models. Meanwhile, GLOC requires the minimum size of
the observation set $M$ to reach $F_1 \geq 0.4$.
}\label{Fig:Syn_F1_CONVP1}
\end{figure*}

\begin{figure*}[t]
\centering
\includegraphics[width=0.95\textwidth]{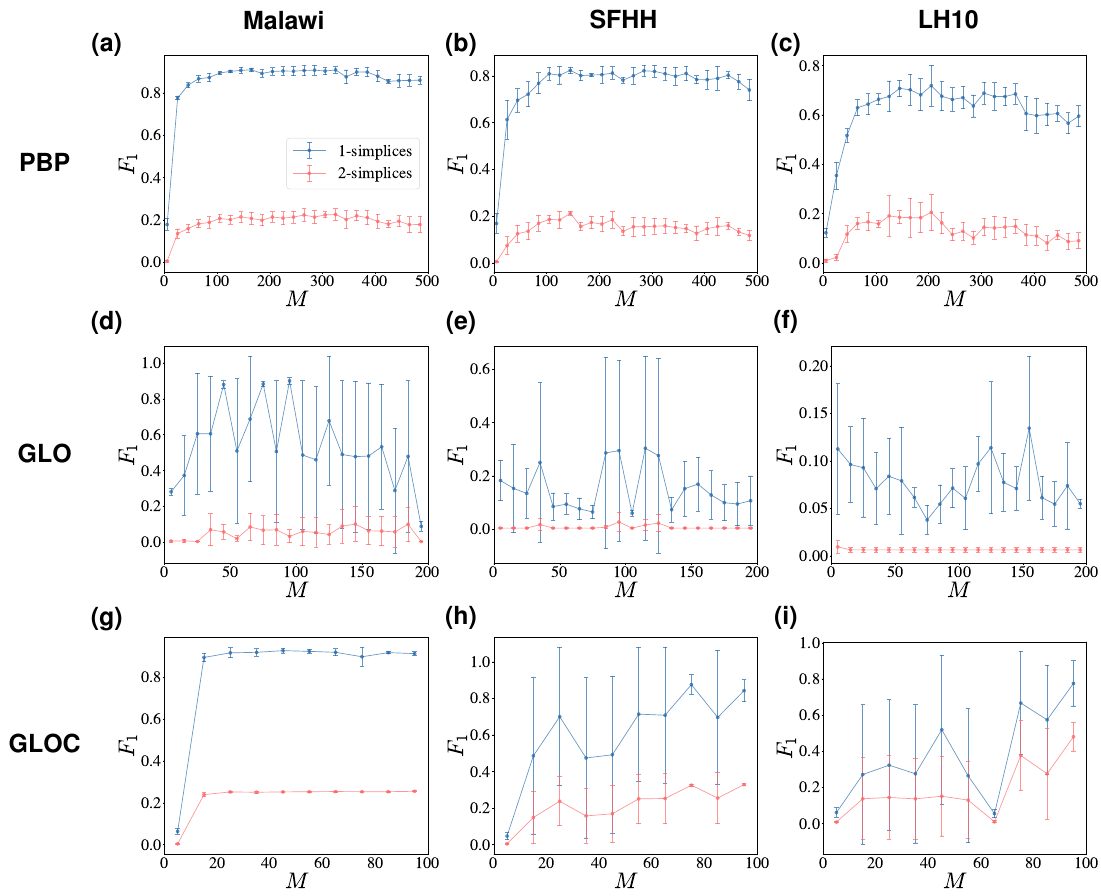}
\caption{\textbf{GLOC is the best performing method to
reconstruct empirical complexes with contaminated strategies.} The 
relative performance of GLO on noisy data ($\psi$ = 1 \%) in strategies
from a Prisoner’s Dilemma game drastically drops with respect to the 
noiseless case (Fig.~4). Meanwhile, compared with 
other two methods, GLOC can still obtain a relatively higher $F_1$ 
score with the smallest observation data sets.
}\label{Fig:Emp_F1_CONVP1}
\end{figure*}